\documentclass[twocolumn]{aastex631}

\usepackage{booktabs}
\usepackage{array}
\usepackage{ragged2e}
\usepackage{graphicx}
\usepackage[flushleft]{threeparttable}

\usepackage{CJK}
\newcommand{\ctext}[1]{\textup{\begin{CJK*}{UTF8}{bkai}#1\ignorespacesafterend\end{CJK*}}}

\usepackage{tabularx}

\begin{document}

\newcommand{\ACC}[1]{\textcolor{red}{#1}}
\newcommand{\AH}[1]{\textcolor{cyan}{#1}}
\newcommand{\RGM}[1]{\textcolor{orange}{#1}}

\title[Dynamical Instabilities in M-dwarf Systems]{Observational Signatures of a Previous Dynamical Instability in Multi-planet M-Dwarf Systems}

\correspondingauthor{Anna Childs}
\email{anna.childs@northwestern.edu}

\author[0000-0002-9343-8612]{Anna C. Childs}
\affiliation{Center for Interdisciplinary Exploration and Research in Astrophysics (CIERA) and Department of Physics and Astronomy Northwestern University,
1800 Sherman Ave, Evanston, IL 60201 USA}

\author[0009-0006-0944-4275]{Alexa P.S. Hua}
\affiliation{Center for Interdisciplinary Exploration and Research in Astrophysics (CIERA) and Department of Physics and Astronomy Northwestern University,
1800 Sherman Ave, Evanston, IL 60201 USA}
\affiliation{Undergraduate student, College of Letters and Science, University of California, Berkeley, 101 Durant Hall, Berkeley, CA 94720}

\author[0000-0003-2401-7168]{Rebecca G. Martin}
\affiliation{Nevada Center for Astrophysics, University of Nevada, Las Vegas,
4505 South Maryland Parkway, Las Vegas, NV 89154, USA}
\affiliation{Department of Physics and Astronomy, University of Nevada, Las Vegas,
4505 South Maryland Parkway, Las Vegas, NV 89154, USA}

\author[0000-0003-2589-5034]{Chao-Chin Yang (\ctext{楊朝欽})}
\affiliation{Department of Physics and Astronomy, The University of Alabama, Box~870324, Tuscaloosa, AL~35487-0324, USA}

\author[0000-0002-3881-9332]{Aaron M. Geller}
\affiliation{Center for Interdisciplinary Exploration and Research in Astrophysics (CIERA) and Department of Physics and Astronomy Northwestern University,
1800 Sherman Ave, Evanston, IL 60201 USA}



\begin{abstract}

We identify observational signatures suggesting a history of dynamical instability in 26 out of 34 M-dwarf multi-planet systems containing no large planets. These systems may have primarily formed in a gas-rich environment, potentially hosted more planets and were more compact. We extend previous simulations of the formation of the TRAPPIST-1 system to $100 \, \rm Myr$ to test the stability of these systems without gas. We find the absence of a strong mean motion resonance in the innermost planet pair and the absence of three body resonances throughout the system are likely to result in the merging and ejection of planets after the gas disk disperses. The runs that experience such an instability tend to produce final systems with lower multiplicities, period ratios larger than two, increased orbital spacings, higher planetary angular momentum deficits, and slightly smaller mass ratios between adjacent planets. Remarkably, we find these same trends in the observations of M-dwarf multi-planet systems containing no large planets. Our work allows us to identify specific systems that may have experienced an instability and suggests that only $\sim25\%$ of these systems formed in their current observed state while most systems were likely more compact and multiplicitous earlier in time. Previous research indicates that planets that have experienced a late stage giant impact may potentially be more habitable than those that did not. With this in mind, we suggest systems around M-dwarfs that contain period ratios larger than two be given priority in the search for habitable worlds.

\end{abstract}

\keywords{Exoplanet dynamics, Exoplanet formation, Extrasolar rocky planets, Exoplanet evolution}


\section{Introduction}

TRAPPIST-1 (T1) is an M-dwarf star that hosts seven terrestrial planets (b-h) that are close in size to Earth \citep{Gillon2016, Gillon2017}. While the planets all orbit within $0.1 \, \rm au$, planets  d, e, and f may have comparable equilibrium temperatures to Earth due to the low luminosity of T1 \citep{Luger2017}. The potential habitability of these planets has led to significant interest in the system, making it one of the most observed and theoretically studied exosystems \citep[e.g][]{Ormel2017, Luger2017, Agol2021, Raymond2022, Huang2022, Grimm2018, Childs2023, Clement2024, Greene2023, VanGrootel2018, Pichierri2024, Ogihara_2009, Lim2023, Coleman2019, Burgasser2017, Lincowski_2018,Schoonenberg_2019}.

T1 is a very compact system, with all planets located at orbital radii in the range $0.01-0.4 \, \rm au$ from the star.  It exhibits one of the longest resonant chains known to date, including two-body Mean Motion Resonances (MMRs), and three-body resonances (3BRs) \citep{Luger2017}. While the outer-most planets (e-h) reside in first-order MMRs with their respective inner adjacent planet, the three inner planets (b-d) have higher-order MMRs (5:3 and 8:5). The 3BRs found in the T1 system are comparable to the Jovian satellites Io, Europa, and Ganymede. Numerous studies have shown that resonant chains are the natural result of inward Type-I planet migration during the planet formation process when the gas disk is still dominant \citep{McNeil_2005, Cresswell_2006, terquem2007, Ogihara_2009, ida2010}. 

One of the leading formation theories for the T1 planets \citep{Ormel2017} suggests that planet formation began at the ice line, where planetesimals first formed by the streaming instability \citep{Johansen2014}. Through  migration, the planetesimals migrated inwards and grew via pebble accretion \citep[see, e.g.,][and references therein]{Johansen2017}. This process results in planet production one at a time where the first planet stalls at the outer edge of the magnetosphere and the subsequent planets stack up behind this planet in resonant locking \citep{Ormel2017, Schoonenberg_2019,Coleman2019, Childs2023}.

Compact planetary systems and resonant chains can be explained by pebble accretion and inward Type-I migration in the absence of outer giant planets \citep{Ogihara_2009,Schoonenberg_2019}.  However, the presence of giant planets induces pressure bumps and gaps in the gas disk which suppresses the influx of pebbles into the inner disk regions thus inhibiting planet growth from pebble accretion \citep{Mulders2021}.  Because of this, it is likely that inner terrestrial systems form primarily through accretion in the presence of outer giant planets which produces more widely spaced orbits, as in the solar system.

Giant planet occurrence rates increase with stellar mass and stellar metallicity \citep{Gonzalez1997, Fischer2005, Ghezzi2010, Wang2015, Ghezzi2018,Childs2022mdwarf}. The relatively low occurrence rates of giant planets around M-dwarfs might suggest that these systems preferentially form terrestrial planets through pebble accretion and thus harbor more high multiplicity systems in resonant chains.  However, observational data do not appear to support this expectation \citep{NASAExoplanetArchive}.  There are many systems that have six planets but only four are around K-type stars (TOI-178, Kepler-80, HD 219314, HD 110067) and the rest are found around more massive stars. If we consider systems with lower planet multiplicities, there are five systems around M-dwarfs with five planets and the number of systems increases as the planet multiplicity decreases.  T1 appears to be an anomaly in the sense that it is the only high-multiplicity system in a compact resonant chain around an M-dwarf.  However, detecting such low-mass planets is difficult, so it is hard to evaluate the prevalence of T1-like systems throughout the Galaxy. 

Pebble accretion is not the only possible formation channel for the T1 planets.  Studies have been able to successfully reproduce the broader characteristics of observed M-dwarf systems by numerically modeling in situ planetesimal accretion in a gas free environment around a small star \citep{Raymond2007, Hoshino2023}, including the T1 system \citep{Coleman2019, Clement2024}. Late stage giant impacts are a common feature of in situ core accretion in a gas free environment\citep{Quintana2016}. As a result, there may be some degeneracy in the observational signatures in systems that form in this way and in the unstable systems we present here. There are valid arguments for and against each formation channel but we focus on pebble accretion as the primary formation channel of the T1 system, and other similar systems, in this paper.

\cite{Childs2023} (henceforth Paper I) numerically modeled the \cite{Ormel2017} formation channel of the T1 planets using an evolving disc model \citep{Hartmann1998} that includes more realistic solid body collisions (ie. fragmentation by \cite{Childs_Steffen_2022}) to simulate the formation of the T1 planets. Their module for {\sc reboundx} also tracks pebble accretion growth \citep{Johansen2017}, Type-I migration, and eccentricity and inclination damping driven by gas \citep{Tanaka_2002}.  They conducted 100 runs starting with moon sized bodies exterior to the ice line.  
They found that most runs resulted in low multiplicity systems of super-Earths which is more representative of the observational data (see Section \ref{sec:Discussion} for further discussion of this) suggesting that forming high-multiplicity compact planetary systems is difficult.

In addition to overcoming difficulties in the formation process, these systems must also remain long term stable after the gas disk dissipates.  The long term stability of both observed exoplanet systems and theoretical systems have been extensively studied \citep{Steffen2013, Tamayo2015, Quarles2017, Rosenthal2019, Wang2018, Rice2023,  Petit2020, Tamayo_2021}.  Theoretical studies of the stability of the observed Kepler systems found that most of these systems are ``dynamically packed" or, filled to capacity \citep{Fang2013F}.  \cite{Volk2020} found that secular chaos can lead to instability over long timescales in nonresonant multiplanet systems and more recently, \cite{Volk_2024} found that small dynamical spacing can be an indicator of instability on a relatively short timescales.  

Motivated by n-body studies of planetary systems in a gas free environment, \cite{Pu2015} and \cite{Volk2015ApJ} proposed that initially, observed systems formed with tighter spacings and higher multiplicities, followed by a series of dynamical instabilities dominated by secular gravitational perturbations between the planets. These instabilities reduce the number of planets and increase their orbital spacings through processes such as planet$-$planet collisions and mergers.   

Additionally, multiple mechanisms for destabilizing resonant chains, such as T1, have been proposed.  \cite{Pichierri2020} found that a secondary resonance, which arises between the synodic frequency and the libration frequencies of the MMR, can amplify libration angles and destabilize a system in time.  Resonant regions may also become chaotic due to the secular eccentricity variations periodically sweeping the chaotic edges of MMRs to wider and narrower period ratios \citep{Tamayo_2021} and instabilities in closely packed systems can also be driven by the overlap of 3-body resonances \citep{Quillen2011, Petit2020, Rath2022}.

Typically, previous dynamical studies of system stability are conducted on the order of Gyr but here we are interested in the shorter term stability of systems that primarily form in the presence of the gas, shortly after the gas disk dissipates. Although the exact role of the gas disk in the formation process is not fully understood and it certainly changes with time, the gas disk likely stabilizes the planets in compact chains as the planets form \citep{Cresswell_2006, Cresswell2008, Tanaka_2002}. 

The runs from Paper I were simulated for only $3 \, \rm Myr$ and in the presence of gas.  In this paper, we test if these systems are stable without gas effects by extending these runs for a total of $100 \, \rm Myr$ of simulation time in the absence of gas.  While we abruptly remove gas from our simulations, \cite{Tamayo2017} showed that the stability of the T1 system is sensitive to the gas dissipation rate.  If the gas is slowly removed, the system adiabatically ends up in the gas-free equilibrium configuration, oscillating with small amplitudes.

This paper aims to better understand the factors that drive instabilities in gas-free systems that formed primarily through pebble accretion and explores the observational signatures of systems that have undergone such instabilities. Although abruptly removing gas can prevent a system from reaching a stable configuration in a gas-free environment, the resulting increase in instability provides a larger data set to probe for observational signatures of a previous instability. Nevertheless, the stability of planetary systems and its observational signatures is a separate topic from the formation of these systems, and hence the systems we draw from Paper I can simply be considered as a convenient pool for our investigations in this work.

In Section \ref{sec:Simulations} we describe our simulations, in Sections \ref{sec:preciting_stability} and \ref{sec:100Myr_analysis} we present the results of our simulations, and compare these results to observations in Section \ref{sec:observations}.  Lastly, we summarize our conclusions and provide a discussion of their implications and suggest directions for future work in Section \ref{sec:Discussion}.

\section{Methodology}\label{sec:Simulations}
Our simulations build on Paper I, which considered 100 simulations based on the formation channel of T1 presented by \cite{Ormel2017}. Their simulations began with 30 moon-sized bodies just outside of $0.1 \, \rm au$, the location of the system's ice line, randomly distributed between $0.1-0.15 \, \rm au$ around a $0.09 \, M_{\odot}$ star.  Their simulations modeled mass growth from pebble accretion, inward Type-I migration, and eccentricity and inclination damping driven by an evolving gas disc.  A planet trap was set in the inner cavity of the disc at $\sim 0.01 \, \rm au$ to prevent the in-fall of planets into the host star, presumably by the magnetospheric truncation of the disk at the inner edge. 


We use the n-body code {\sc rebound} and extend all the 100 runs to assess their stability in the absence of gas and compare the resulting systems to observations.  Thus, the only forces being modeled are gravitational in these new simulations. We resolve all collisions as inelastic mergers.  We deviate from Paper I and do not include the effects of fragmentation since we do not include gas effects or pebble accretion, and \cite{Quintana2016} found that in a gas-free environment, fragmentation results in similar systems compared to simulations without fragmentation. However, the accretion timescale doubles if fragmentation is included.  We also deviate from Paper I in how we define a planet.  We now define a planet as a body that has a minimum mass of at least $0.07 \, M_{\oplus}$.  We do this so we may make a more direct comparison to the architectures of the observed systems in Section \ref{sec:observations} and the lowest mass exoplanet observed to date has a mass of $0.07 \, M_{\oplus}$ \citep{NASAExoplanetArchive}.

\section{Predictors of Stability}\label{sec:preciting_stability}
We break the simulated planetary systems into three groups based on their evolution and final planet multiplicity at $t=100 \, \rm Myr$.
T1-analogs (T1A) are runs that form at least six planets in the presence of gas (using our lower mass limit of $0.07 \, M_{\oplus}$) and retain all planets after $100 \, \rm Myr$ of integration time.  The  second group is the primordial low multiplicity systems (PLM) and are the systems that initially formed with fewer than six planets, did not experience any further instabilities after $100 \, \rm Myr$ of simulation time and thus retain their original multiplicity.

Unstable systems (U) are systems that experience an instability in the absence of gas, resulting in a final system with fewer planets than it had initially formed in the presence of gas at $t=3 \, \rm Myr$. These planets may be lost through either ejections or mergers although because the escape speed from these planets is typically less than their orbital velocities, interplanetary collisions are much more common than ejections \citep[e.g.][]{Petrovich_2014}. However, we track both and observe ejections in five out of our 100 simulations. All but one of these five runs also experience a merging of a planet pair as well, and so late stage giant impacts between planets is a common feature in the unstable runs.  This result has implications for the final planet properties as discussed further in Section \ref{sec:Discussion}.  

  Table \ref{tab:run_descripts} lists these three groups and the number of simulations in each group.  We also list the number of runs at $3 \, \rm Myr$ that had six or more planets and those that did not.  The number of runs at $3 \, \rm Myr$ in the T1A row are the number of runs that had at least six planets at that time.  Similarly, the number of runs at $3 \, \rm Myr$ in the PLM row are the number of runs that had less than six planets at that time.

\begin{table*}
\centering
    \begin{tabular}{llcc}
        \hline
         Definition & Group Name & \# Runs & \# Runs  \\
      & &   ($3 \, \mathrm{Myr}$) & ($100 \, \mathrm{Myr}$) \\
        \hline
         \footnotesize TRAPPIST-1 analogs form with at least six planets and retain all planets after gas dissipation & T1A &28 & 9 \\
        \footnotesize Primordial low multiplicity systems form with less than six planets and remain stable & PLM &  72 & 40 \\
        Unstable runs experience at least one instability which decreases the final number of planets & U  &  -  & 51 \\
        \hline
    \end{tabular}
    \caption{The three types of simulation outcomes. Out of our 100 simulations, 28 formed at least six planets, and 72 formed fewer than six at $3 \, \rm Myr$. We classify our simulations based on their final outcomes at $100 \rm \, Myr$, e.g. both categories could go unstable and contribute to U, and the stable ones go into T1A and PLM respectively.}
    \label{tab:run_descripts}
\end{table*}

We have classified the planetary systems accordingly to see if there are differences between stable systems that form early on while the gas is present (either PLMs or T1As), and systems that experience an instability later in time without the stabilizing effects of the gas.  In this section we present our analysis of these three distinct groups of simulations at $t=3 \, \rm Myr$ to better understand what determines the stability of a planetary system after the gas disk dissipates.   

When not undergoing any dynamical instabilities, the final T1A and PLM runs largely have the same system architectures as they did at $t=3 \, \rm Myr$.  After the dissipation of the gas, systems are about equally likely to either experience an instability that results in the loss of a planet or retain the planet multiplicity and mass distribution they had at $3 \, \rm Myr$ in the presence of gas.  Only 9\% of the initial 100 runs are able to form T1As that remain stable after the dissipation of gas.  More common however, 40\% of the initial 100 runs form with a PLM and remain largely unchanged after the dissipation of the gas disk.  


\begin{figure*}
\includegraphics[width=0.5\textwidth]{ 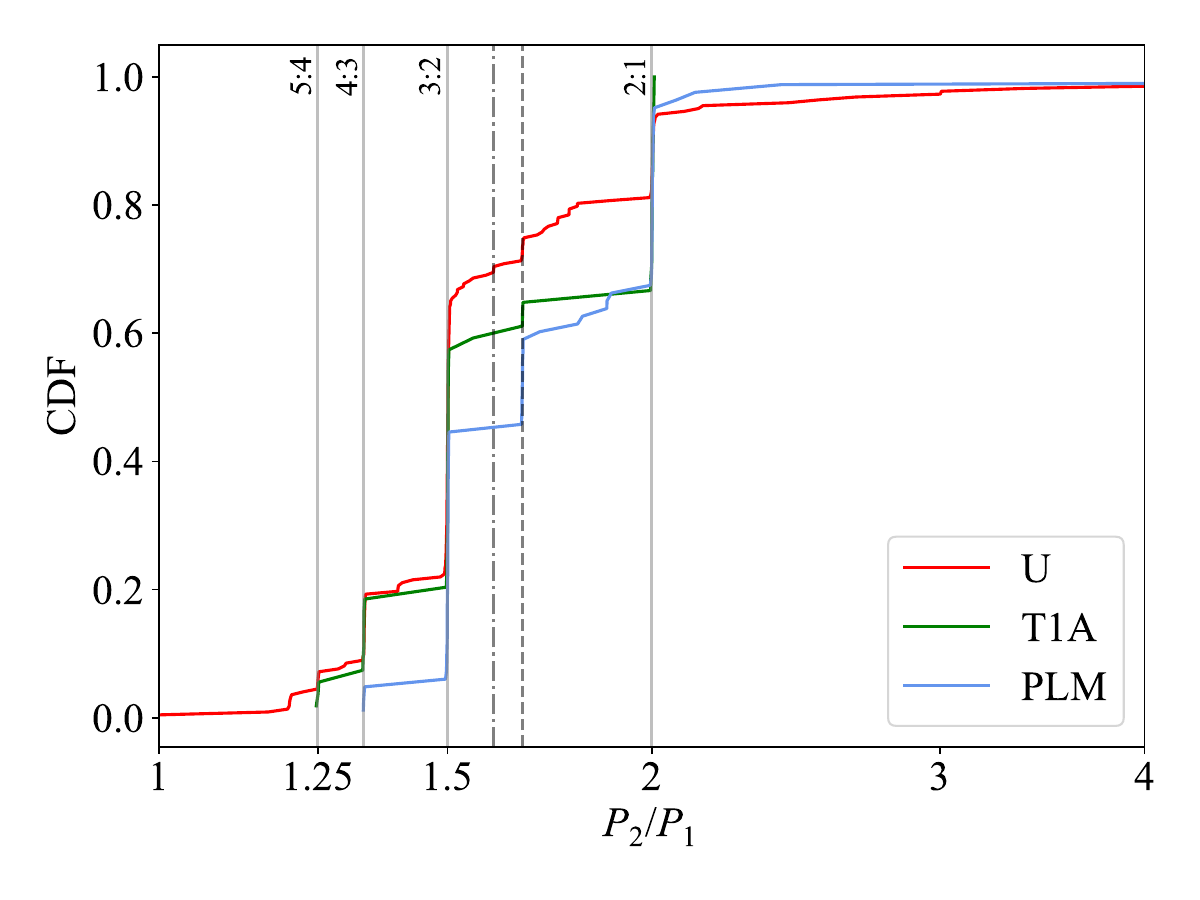}
\includegraphics[width=0.5\textwidth]{ 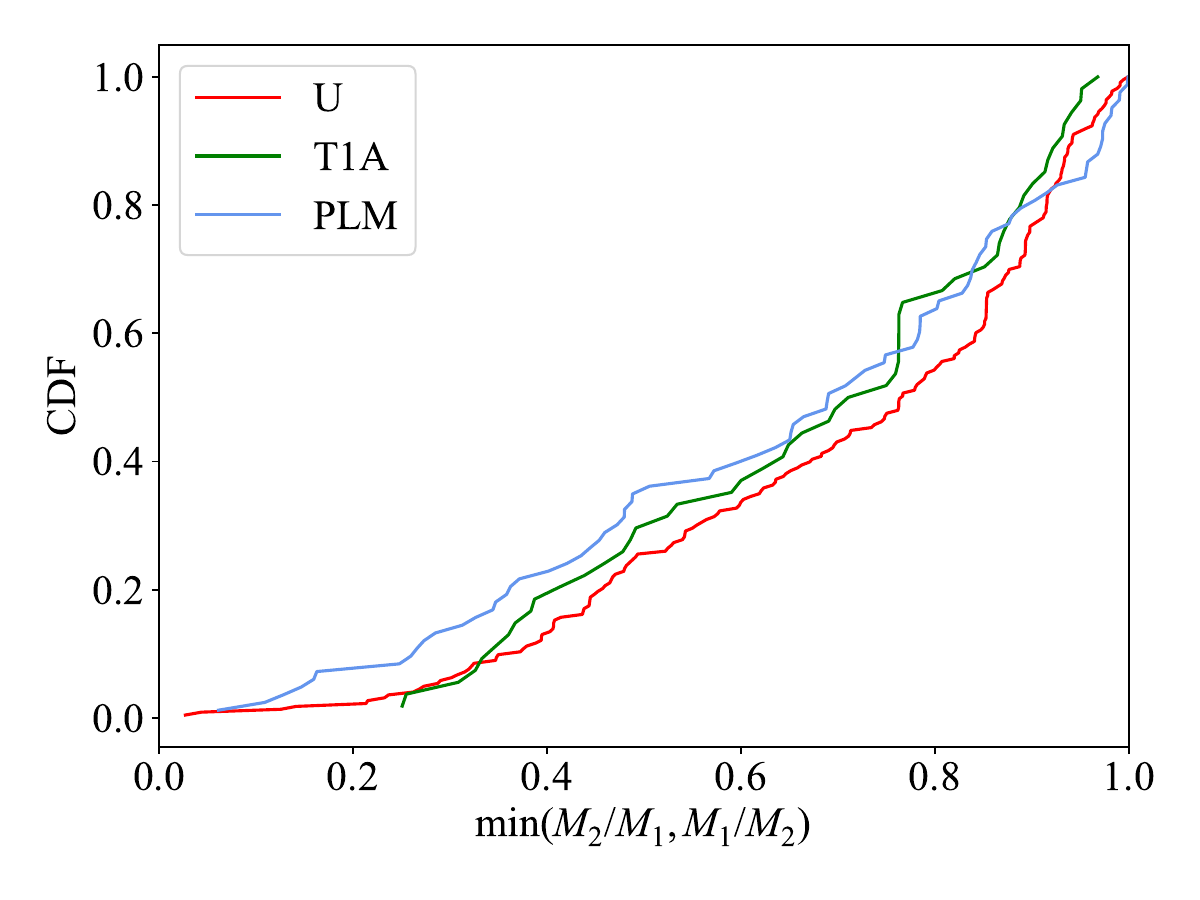}
\includegraphics[width=0.5\textwidth]{ 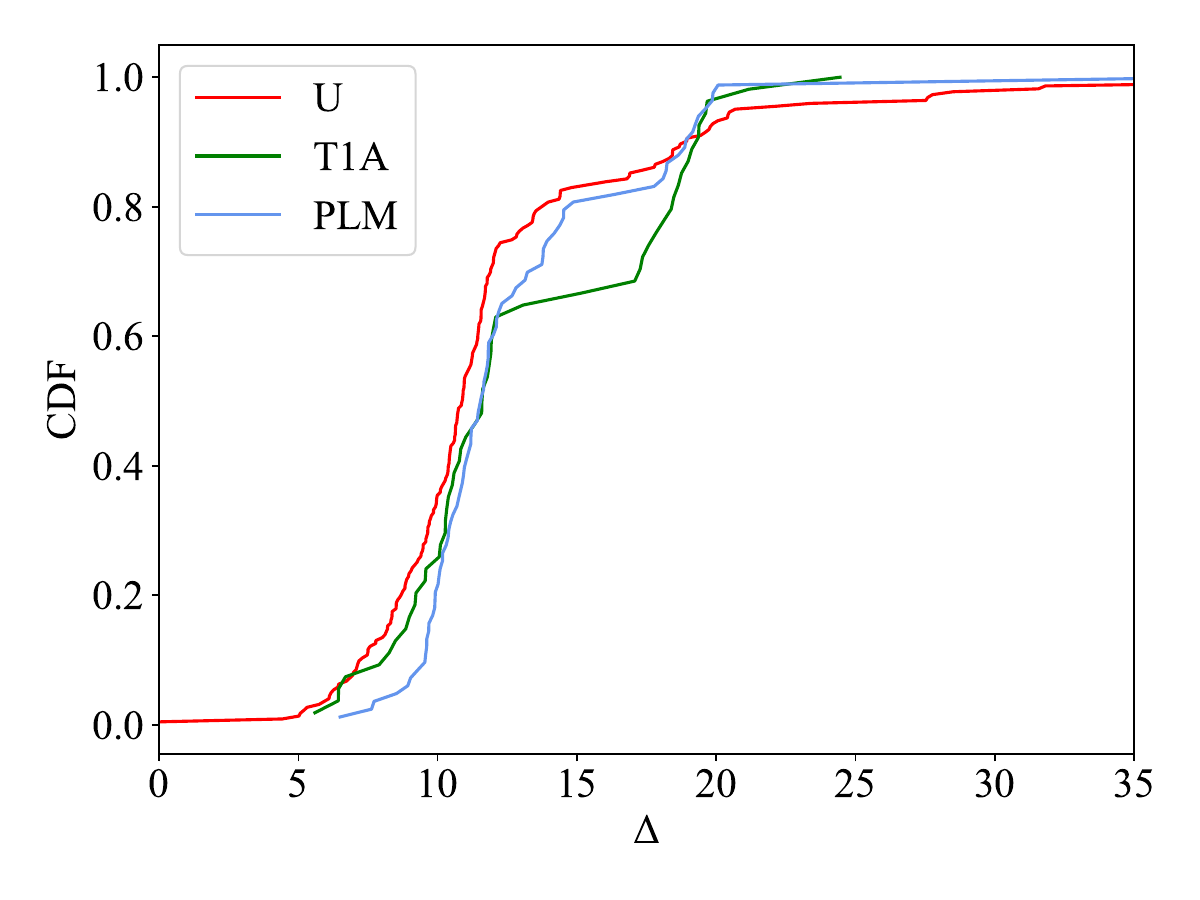}
\includegraphics[width=0.5\textwidth]{ 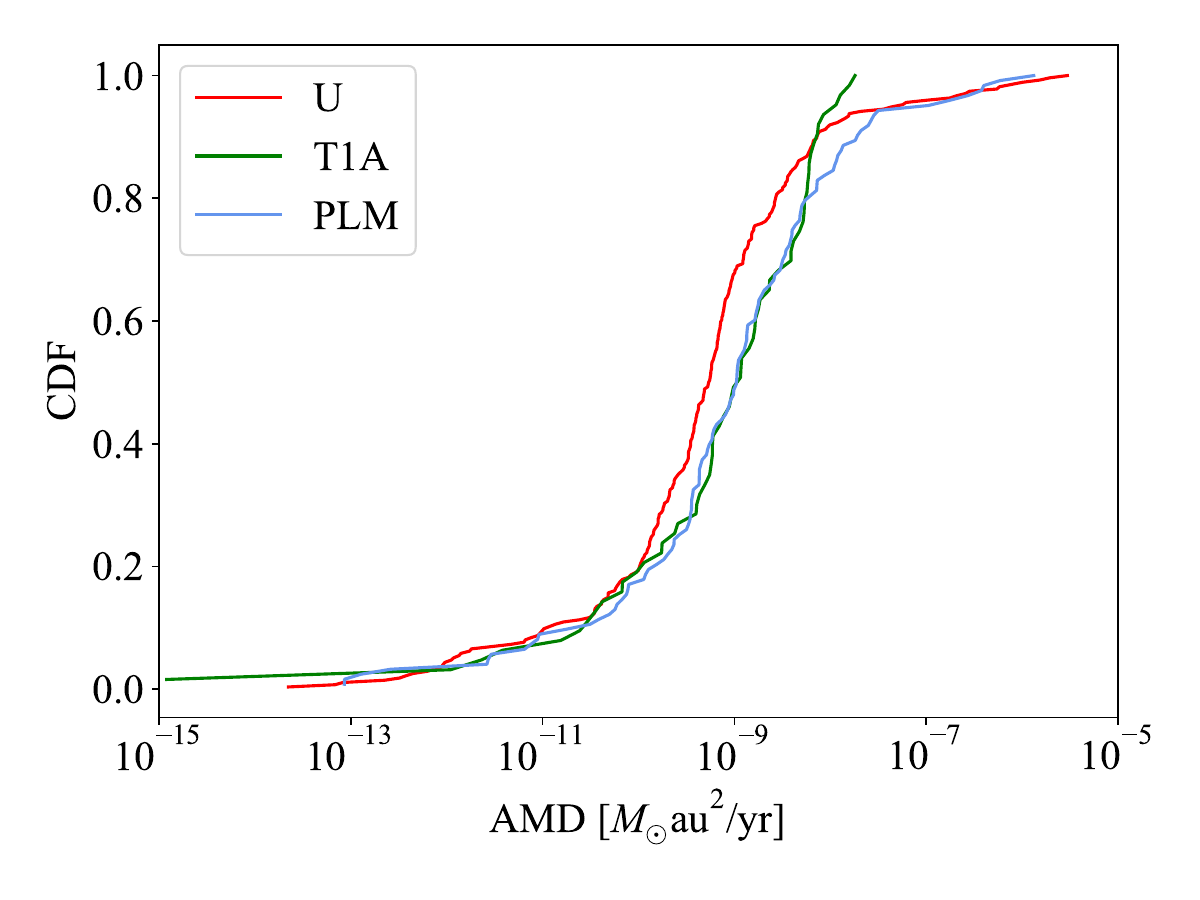}
    \caption{Data for simulated planets at $t=3 \, \rm Myr$ (gas is present), classified by the three different groups at t = 100 Myr listed in Table \ref{tab:run_descripts}. Top left: CDF of period ratios for adjacent planets. The strongest first order MMRs are marked with vertical lines, the 5:3 MMR is marked with a dashed line, and the 8:5 MMR is shown with a dot-dashed gray line.  Bottom left: CDF of orbital spacing between adjacent planets in units of mutual hill radii.  Top right:  CDF of mass ratio for adjacent planets.  Bottom right:  CDF of AMD for each planet.}
    \label{fig:start_CDFs}
\end{figure*}

\subsection{Common stability metrics}\label{sybsec:stability_metrics}
Figure \ref{fig:start_CDFs} shows cumulative distribution functions (CDFs) for four common measurements used to evaluate dynamical stability at $t = 3 \, \rm Myr$ of the three groups classified by $t = \, \rm 100 Myr$.  We also perform K-S tests between all three groups in each measurement.  We consider a $p-\rm{value} = 0.05$ to be the threshold, above which we cannot reject the null hypothesis that the data are from the same distribution.  

In the top left panel is the CDF of the period ratios for the adjacent planet pairs in each group.  The CDFs show a number of 'jumps' (vertical lines).  These jumps coincide with mean motion resonance (MMR) locations and is the expected outcome from inward Type-I migration.  We show the location of the strongest MMRs less than two with gray vertical lines.  These MMRs are all first order MMRs.  The higher-order 5:3 MMR, is shown with a dashed gray line, and the 8:5 MMR is shown with a dot-dashed gray line.  These higher-ordered MMRs are included since they are found in the T1 system. All three groups display a large fraction of planet pairs found near the 3:2 and 2:1 MMRs.  The T1A and PLM planets are almost entirely found near the marked MMRs except the 8:5. By contrast, the initial conditions that undergo instability after disk dissipation (U systems) exhibit more period ratios that are not near significant MMRs: 15\% of period ratios are not within 0.01 of an MMR. This suggests that orbital periods that are not near a strong MMR tend to lead to an instability in the system after gas disk dissipation.  We discuss this further in Section \ref{subsec:3BR}.  K-S tests reveal that we cannot reject the null hypothesis between any pairs of the groups except for the PLM and U period ratios which have a $p$-value$\, \approx 0.0003$.

The bottom left panel shows the CDF of $\Delta$, the orbital spacing between adjacent planet pairs in units of mutual Hill radii,
\begin{equation}
\Delta = \frac{2|a_2 - a_1|}{a_1 + a_2} \left(\frac{3M_\star}{m_1 + m_2}\right)^{1/3},
\end{equation} where $a_2, a_1$ and $m_2, m_1$ are the semi-major axes and masses of the planets, and $M_{\rm \star}$ is the mass of the star.  The U runs display smaller dynamical spacings than the stable T1A and PLM systems.  This is in agreement with \cite{Volk_2024} who found that smaller dynamical spacings are an indicator of instability.  Again, K-S tests reveal that we cannot reject the null hypothesis between any pairs of the groups except perhaps marginally for the PLM and U period ratios which have a $p$-value$\,\approx 0.02$.

The top right panel shows the mass ratio of adjacent planet pairs found in each of the three groups.  Following \cite{Volk_2024}, we take the minimum of the outer to inner and inner to outer mass ratios, so the value is always 1 or less.  The U runs display larger mass ratios than the PLM and T1A runs.  K-S tests reveal that we cannot reject the null hypothesis between any pairs of the groups at this point.

In the bottom right panel we show the CDF for the planet angular momentum deficit (AMD) in each group.  The AMD for a planet in a system is calculated by
\begin{equation}
\mathrm{AMD} =  m \sqrt{G M_\star a} \left(1 - \sqrt{1 - e^2} \cos i \right),
\end{equation} 
where $m, a,e,$ and $i$ are the mass, semi-major axis, eccentricity and inclination relative to the initial orbital plane of the planet.  A smaller AMD indicates the orbit is nearly circular and coplanar to the initial orbital plane while a larger AMD indicates a more excited orbit.  Counterintuitively, we see that the U planets have a lower AMD than the PLM and T1A planets, which have a larger and similar planetary AMD.  We can reject the null hypothesis between the U data and both the PLM and T1A data ($p-\rm{value}=0.001$ and 0.02, respectively) but not between the PLM and T1A data.

The relatively low AMD of the U planets may be attributed to lower planet eccentricities in the U runs.  All runs, with an exception of a few runs in the PLM and U systems, show a zero inclination due to the efficient gas damping.  There is more variation in eccentricity values however, with the U planets displaying the lowest eccentricity values.  We find that the U systems that do not contain all MMRs have low median planet eccentricities ($e \approx 0.005$) as planet pairs in non-resonant period ratios in a gaseous disk tend to approach a zero eccentricity in contrast to those in resonances which tend to reach a finite eccentricity \citep[e.g.][]{Goldreich2014}.


\begin{figure}
\includegraphics[width=0.5\textwidth]{ 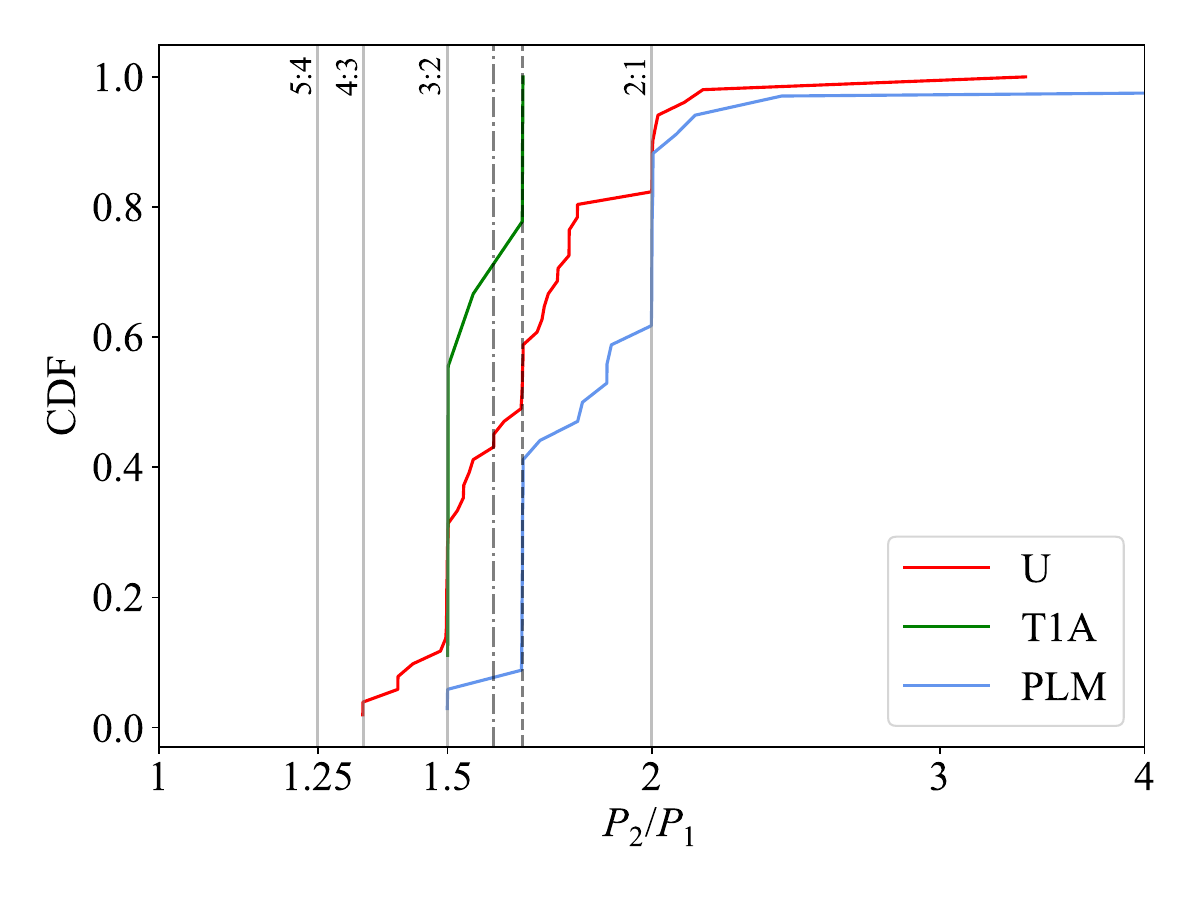}
\includegraphics[width=0.5\textwidth]{ 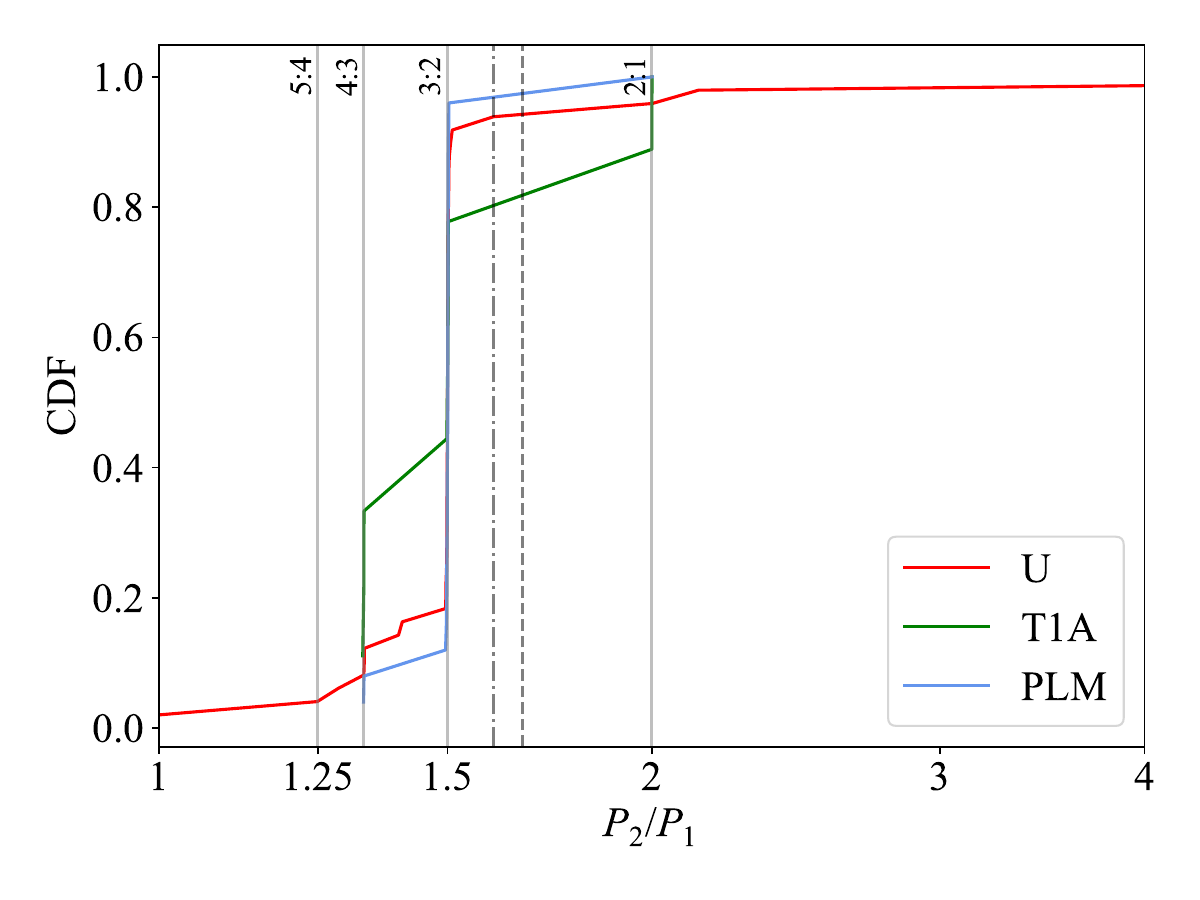}
\includegraphics[width=0.5\textwidth]{ 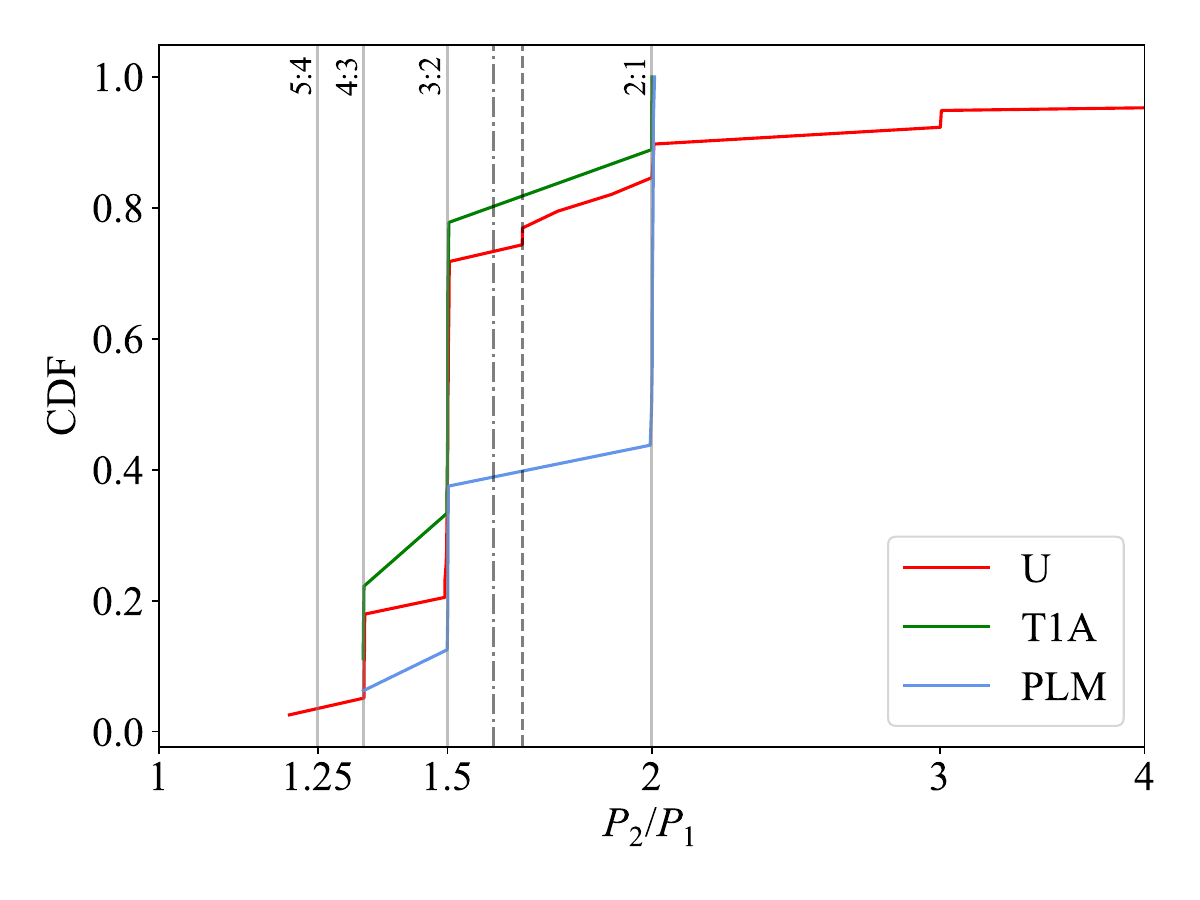}
    \caption{CDFs of the period ratios for the first planet pair (top), second planet pair (middle) and third planet pair (bottom) at $3 \, \rm Myr$.  The strongest first order MMRs are marked with vertical lines, the 5:3 MMR is marked with a dashed line, and the 8:5 MMR is shown with a dot-dashed gray line.}
    \label{fig:PP_MMRs}
\end{figure}
\subsection{Planet pair period ratios and 3BRs}\label{subsec:3BR}

We take a closer look at the distributions of the period ratios of the adjacent planet pairs throughout the systems. Specifically, we are interested in better understanding where the strong MMRs exist in each system.  Figure \ref{fig:PP_MMRs} shows the CDF for period ratios for the first, second, and third planet pairs in the systems.  Again, the strongest first order MMRs are marked with vertical lines, the 5:3 MMR is marked with a dashed line, and the 8:5 MMR is shown with a dot-dashed gray line.

The second and third planet pairs are almost exclusively found in a first order MMR, in all groups.  The first planet pair displays more variation in period ratios however.  The first planet pair in the T1A runs are almost always found in either a 3:2 or a 5:3 MMR.  The first planet pairs in the PLM runs show more variation.  The PLM runs include 2:1 MMRs and also period ratios that are not near a strong first order MMR.  By contrast, the U runs display the largest variation of period ratios for the first planet pairs.  Most period ratios in the U runs are not found near an MMR, indicating that it may be the period ratio of the innermost planet pair that affects the stability of the system after gas disk dissipation.

We also check our systems for the presence of three body MMRs, where the inner planet pair and the outer planet pair of an adjacent triplet are both in a MMR.  If we consider an adjacent planet triplet with mean motions $n_0, n_1,$ and $n_2$, a 3BR satisfies the condition
\begin{equation}
    k_0 n_0 +  k_1 n_1 +  k_2 n_2 \approx 0,
\end{equation}
where $k_0,k_1,$ and $k_2$ are small integers.  The order of the 3BR is defined as $q=|k_0+k_1+k_2|$.  The lower the order of the 3BR, the stronger the 3BR with the zeroth order being the strongest \citep[e.g.][]{Nesvorny1998, Gallardo2016}.

\begin{figure}
\includegraphics[width=.5\textwidth]{ 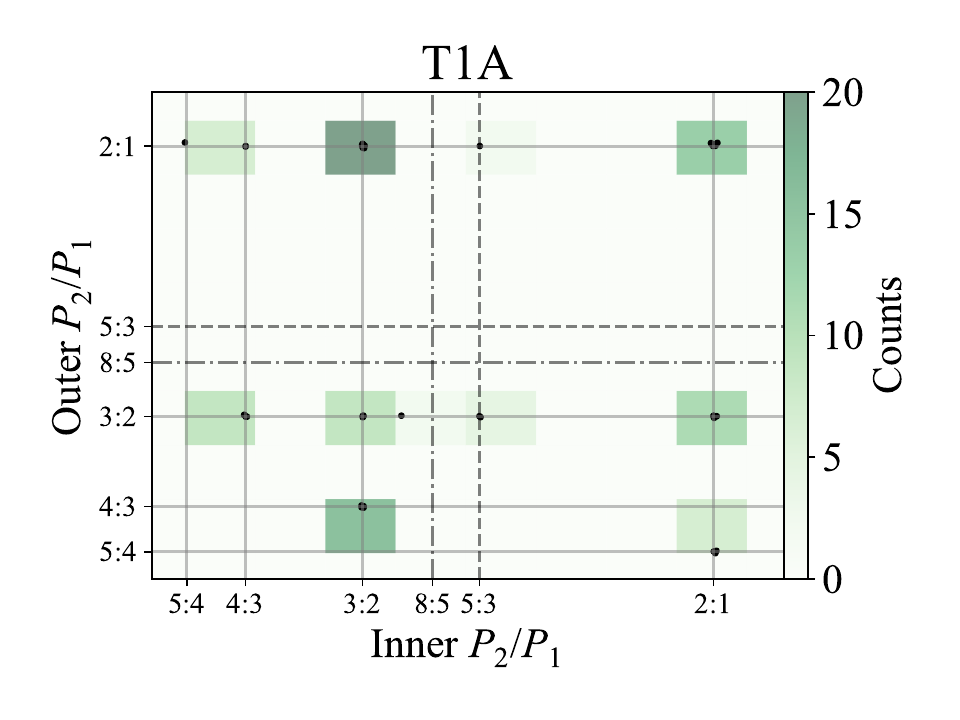}
\includegraphics[width=.5\textwidth]{ 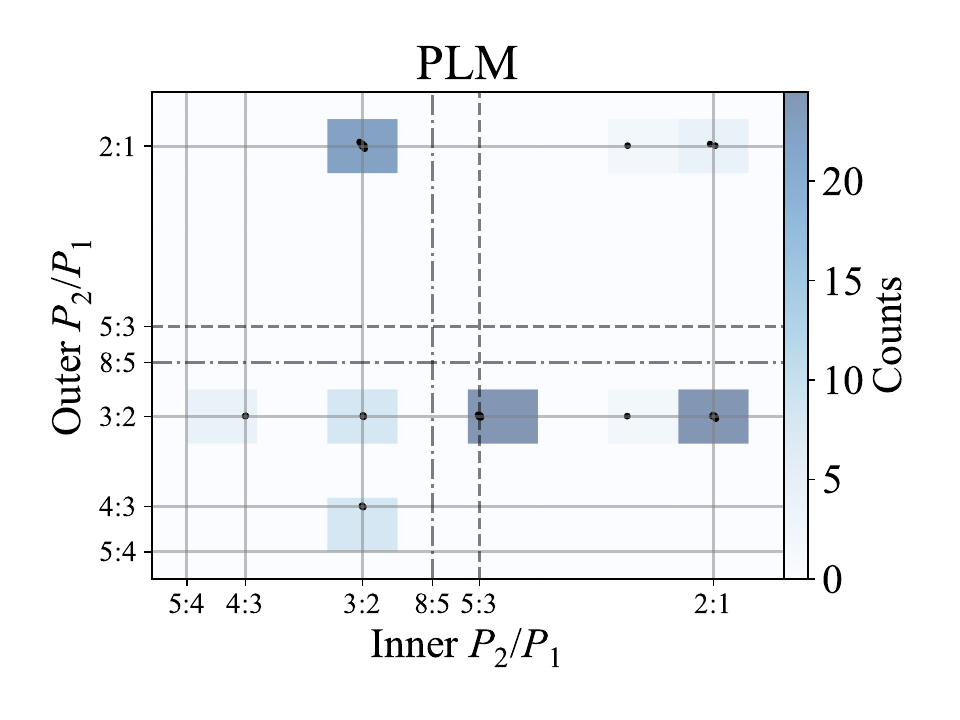}
\includegraphics[width=.5\textwidth]{ 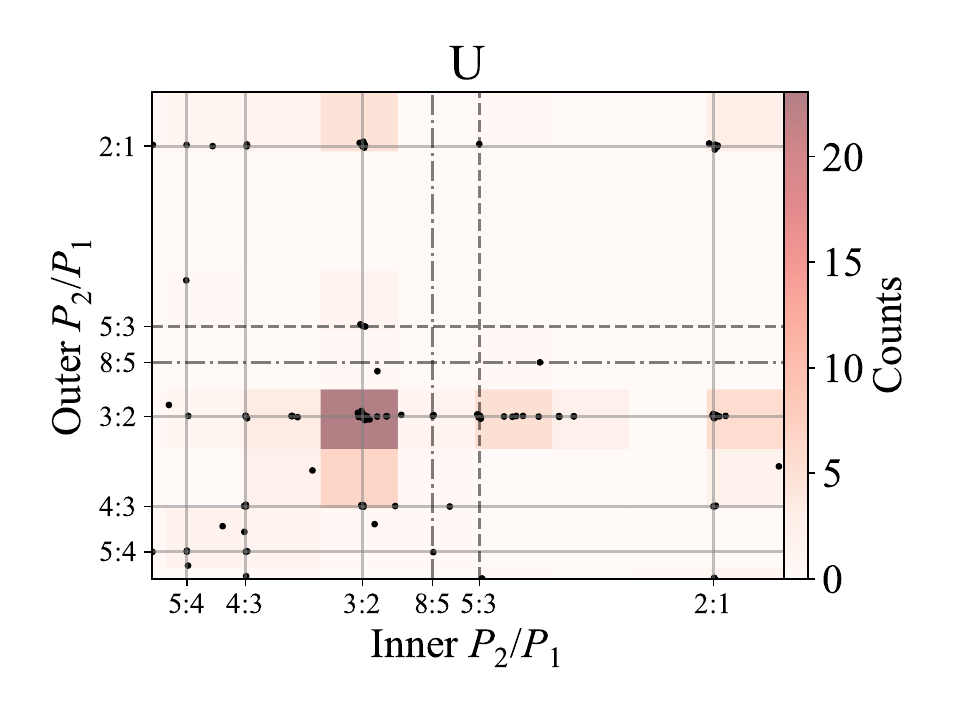}
    \caption{2D histograms and scatter plots of the period ratios for each adjacent planet triplet in each group at $3 \, \rm Myr$.  The period ratio of the inner planet pair is on the horizontal axis and the period ratio of the outer planet pair is on the vertical axis.  The strongest first order MMRs are shown with gray solid lines, the 5:3 MMR is shown with a dashed line, and the 8:5 MMR is shown with a dot-dashed gray line.}
    \label{fig:3BRs}
\end{figure}

Figure \ref{fig:3BRs} shows the period ratios of the inner planet pair on the horizontal axis and the period ratio of the outer planet pair for the adjacent planet triplets in each group of runs.  We consider all possible adjacent planet triplets in the system.  For example, the first planet triplet includes the first, second, and third planets, the second planet triplet includes the second, third, and fourth planets, and so on.  The first order MMRs are marked with solid gray lines, the 5:3 MMR is marked with a dashed gray line, and the 8:5 MMR is shown with a dot-dashed gray line.  All planet triplets except for one, in the T1A are found in 3BRs.  Furthermore, these 3BRs are all mostly zeroth as the MMRs of the inner and outer planet pairs are all first order.  The 3BRs that have a planet pair in the 5:3 MMR are first order 3BRs.  In the PLM runs we find a similar trend.  The vast majority of planet triplets are found in strong zeroth or first order 3BRs with the exception of two triplets.  

In contrast to the PLM and T1A runs, the planet triplets in the U runs are largely found outside of prominent 3BRs.  Some of the planet period ratios extend to very large values but we only show the parameter space of interest.  While at least one of the planet pairs in a triplet may be found near a MMR, strong 3BRs are largely absent in the U runs.  This finding suggests that strong three body mean motion resonances may be needed to maintain the original planetary system after the dissipation of gas and become increasingly important with planet multiplicity.

\section{Signatures of a Dynamically Unstable History}\label{sec:100Myr_analysis}

\begin{figure*}
\includegraphics[width=0.5\textwidth]{ 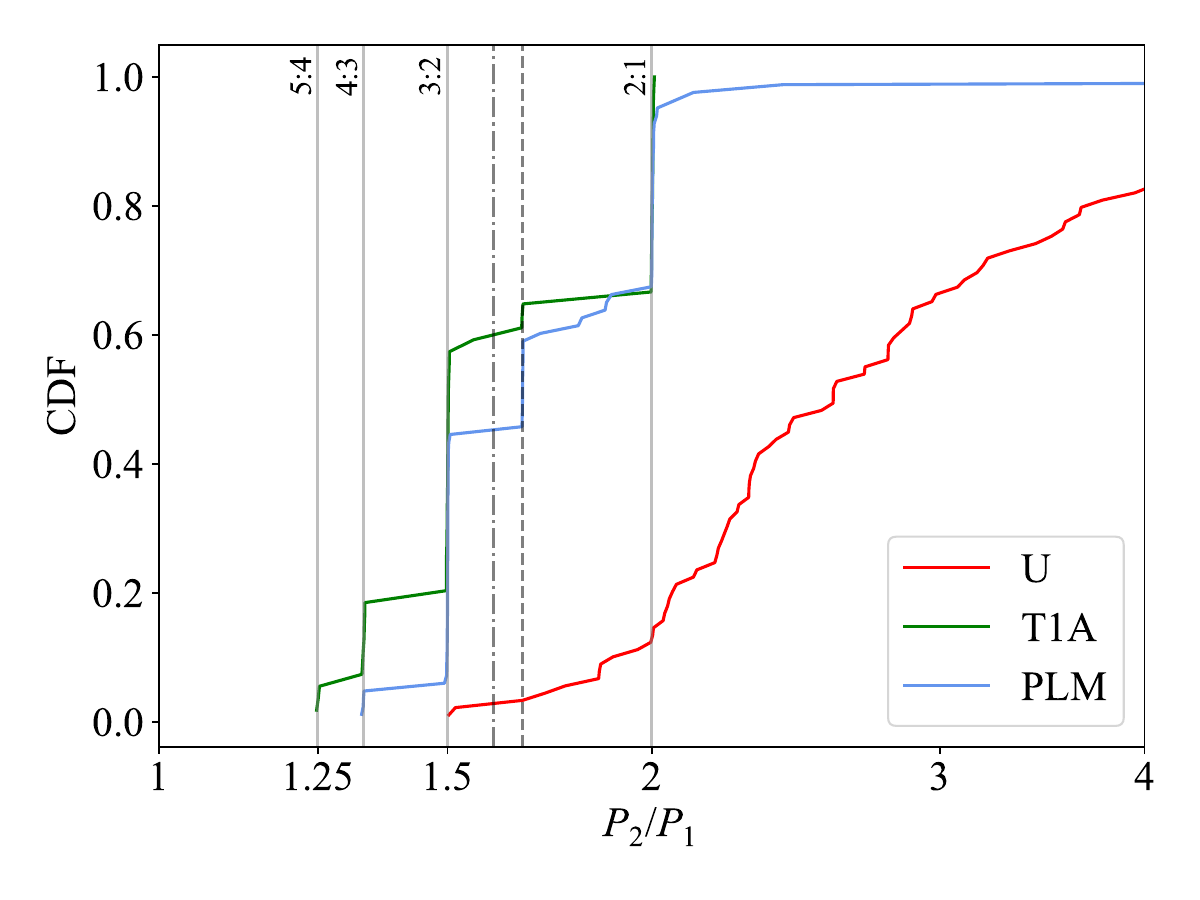}
\includegraphics[width=0.5\textwidth]{ 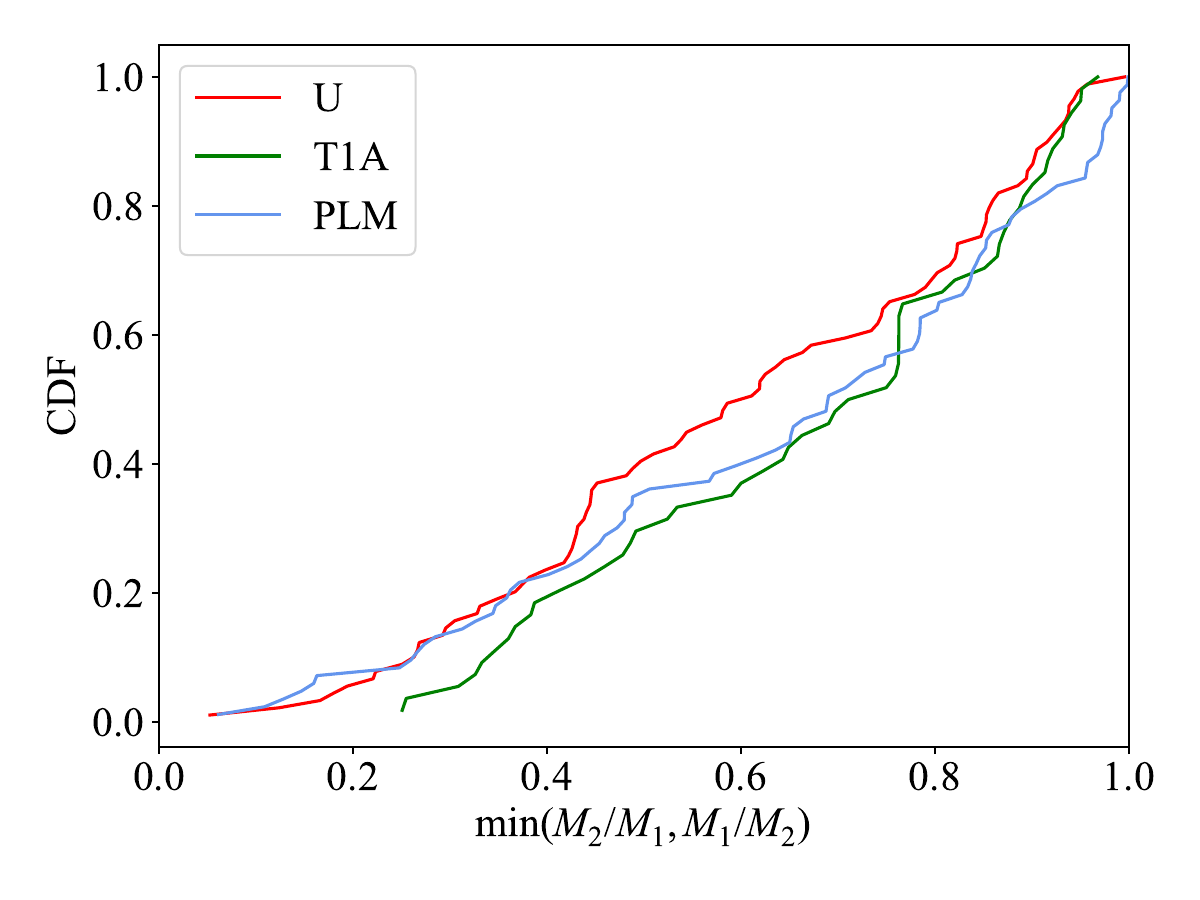}
\includegraphics[width=0.5\textwidth]{ 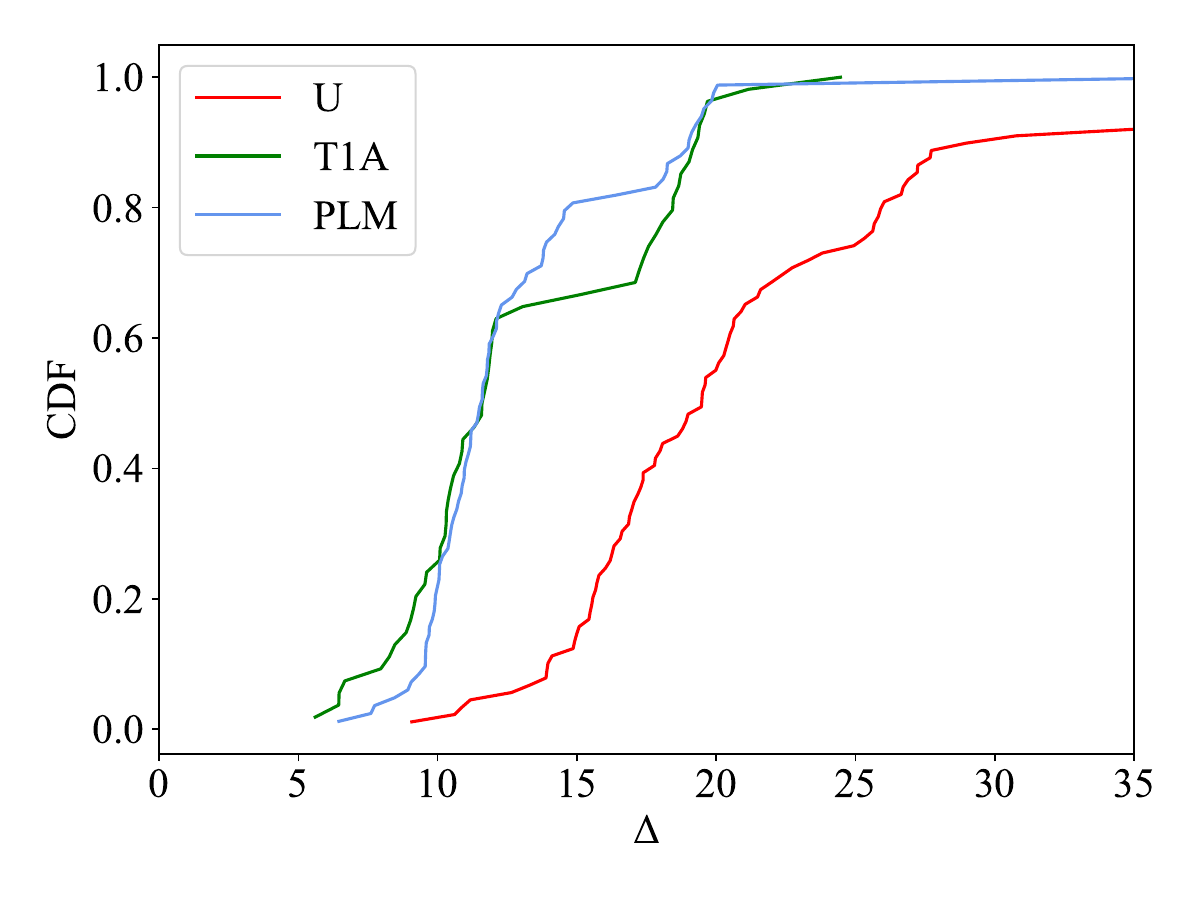}
\includegraphics[width=0.5\textwidth]{ 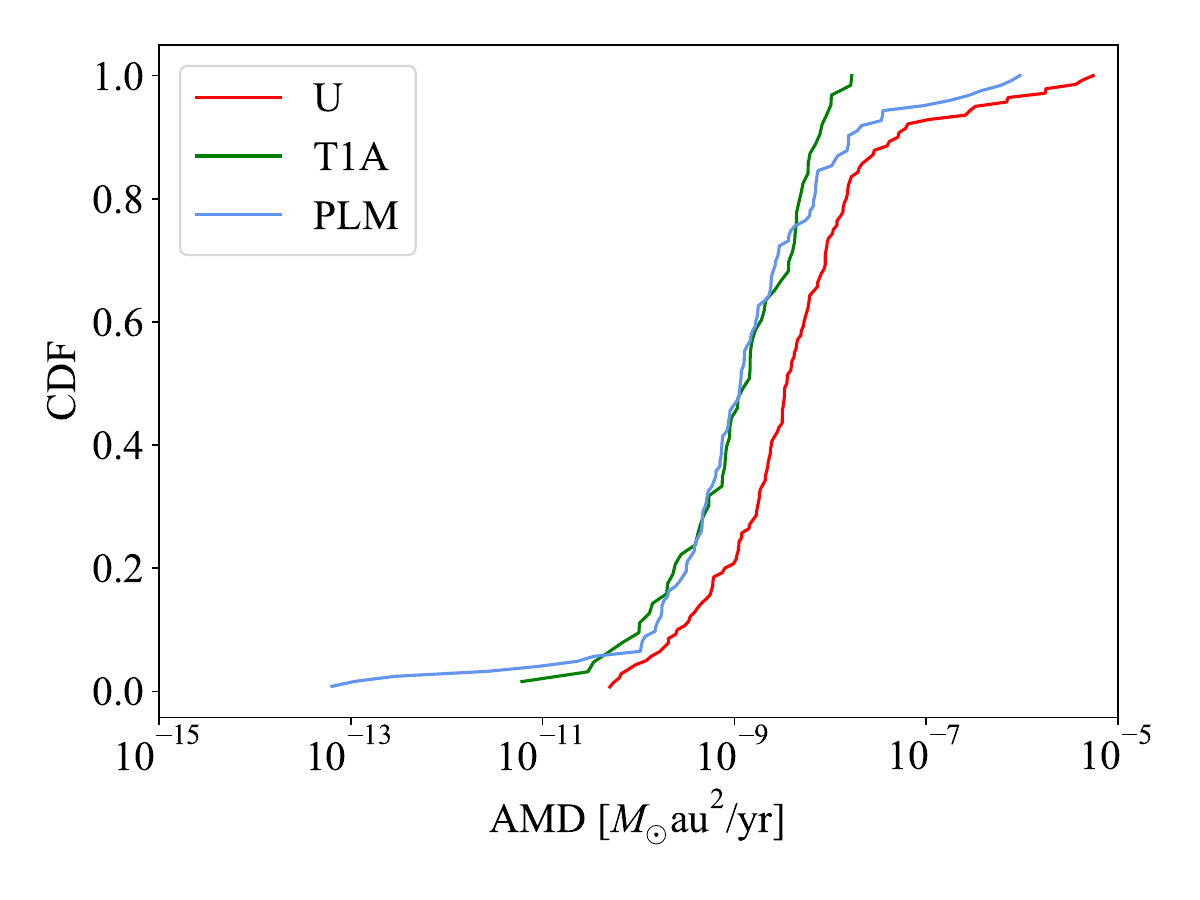}
    \caption{Data for simulated planets at $t=100 \, \rm Myr$ (no gas is present). Top left: CDF of period ratios for adjacent planets. The strongest first order MMRs are marked with vertical lines, the 5:3 MMR is marked with a dashed line, and the 8:5 MMR is shown with a dot-dashed gray line.  Bottom left: CDF of orbital spacing between adjacent planets in units of mutual hill radii.  Top right:  CDF of mass ratio for adjacent planets.  Bottom right:  CDF of AMD for each planet.}
    \label{fig:end_CDFs}
\end{figure*}

We now consider the system architectures after $100 \, \rm Myr$ of integration time in a gas-free environment.
Figure \ref{fig:end_CDFs} is the same as Figure \ref{fig:start_CDFs} except the data are for the final systems that have been integrated to $100 \, \rm Myrs$ in the absence of gas.  The T1A and PLM experience little evolution and so the distributions of the stability metrics of these systems are nearly the same as they were at $3 \, \rm Myr$ in the presence of gas.  On the other hand, the U systems experience significant dynamical evolution that results in at least two planets merging together or one planet being ejected.

The final U systems now have system architectures with larger orbital spacings, period ratios, and planetary AMDs.  This is to be expected from instabilities that result in planet mergers and scattering.  Similarly, we see that the mass ratios in the U runs are shifted to lower values indicating the systems now have less equal massed planets than before.

In addition to the final U systems having significantly different architectures from their original configuration in the presence of gas, we also note that the period ratio, AMD, and $\Delta$ distributions are now statistically different from those of the PLM and T1A runs by finding K-S test $p-$values $\ll 0.05$ for all three metrics.  We do not find a statistical difference between the three groups of the mass ratios for adjacent planets but note that the U mass ratios are now smaller on average, having an average planet mass ratio of 0.6 compared to 0.65 and 0.67 of the PLM and T1A runs, respectively.

Yet another and arguably the most important distinction between the U and the other two groups is the period ratio distributions.  While the period ratios of the T1A and PLM runs are almost exclusively found near MMRs less than two, we see that 88\% of the U period ratios are found at values greater than two. 

Lastly, when considering the final multiplicities of the systems we find notable differences between all three sets.  Of course, we expect systems that experience an instability in the form of a merger to have a lower system multiplicity and for the T1A systems to have the largest median multiplicity as our definition of a T1A is a system that has at least six planets.  When comparing the PLM and U multiplicities we find the distributions are statistically indistinguishable. However, U systems have a median planet multiplicity of two and the PLM runs have a median planet multiplicity of three.  

These findings suggest observable differences between systems that underwent an instability, likely involving a planet merger or ejection after the dissipation of gas, and systems that remained stable, retaining their original planetary configuration formed in the presence of gas. In order to test this, we next consider the observations.

\section{Observational Tests}\label{sec:observations}
Motivated by the results of our simulations, we hypothesize that there exist two distinct populations in the observed exoplanetary systems that initially formed in a similar manner as the T1 system: a population that experienced an instability, most likely in the form of a planet merger or scattering event after the dissipation of gas, and a population that did not.  Our simulations show that multi-planet systems that experience an instability after gas dissipation are more likely to have period ratios greater than two with larger $\Delta$ and AMD values than the systems that have period ratios less than two and are found near MMRs.  Additionally, systems that experienced an instability should have a smaller adjacent planet mass ratio than those systems that did not.

 \subsection{Methodology}
Our observational data comes from \cite{NASAExoplanetArchive}.  We are interested in systems that may have formed in a similar way as T1 and so we only consider multi-planet systems around M-dwarfs ($M_{\rm \star}\leq 0.6 \, M_{\odot}$ or contains "M" in the listed spectral type) that do not host a planet with a mass greater than $10 \, M_{\oplus}$ or a radius larger than $2 \, R_{\oplus}$.  We exclude systems with planets outside of the Earth and Super-Earth regime as more massive planets can influence the formation of the terrestrial system in ways not explored here.  There are 34 systems with 92 planets that meet this criteria.  

To test our hypothesis we separate these 34 systems into two groups: systems that contain at least one period ratio greater than two and those that do not.  We refer to the group of systems that contain a period ratio greater than two as `Big PR' and the group of systems that do not as `Small PR'.  We break the observations up accordingly since we found from the simulations that the most notable distinction between the systems that experienced an instability and those that did not was the presence of period ratios larger than two.  We find that $\sim75\%$ of the exoplanet systems in our sample contain at least one period ratio greater than two.  There are 64 planets in 26 systems in the Big PR group and 28 planets in 8 systems in the Small PR group.

While all planets have measurements for their orbital period, many planets lack mass measurements but instead have radius measurements.  Because some of our previously considered metrics for stability depend on the planet mass, we use the probabilistic mass-radius relationship model in the MCMC code FORECASTER \citep{Chen2017} with the observed planet radii to approximate the planet mass.  We use these randomly sampled masses when there are no planet mass or $m\mathrm{sin}i$ measurements available.  

To assign a planetary mass, therefore, we first use the observed planet mass if available.  If not, we use the observed planet $m$sin$i$ (assuming an inclination of $90^{\circ}$), and if neither are available, we use the observed planet radius with the upper error on the radius measurement to get the approximate planet mass from FORECASTER.  We list all the Small PR systems in Table \ref{tab:SMALL_PR_obs_data} and all the Big PR systems in Table \ref{tab:BIG_PR_obs_data}.

We note that 53\% of the planets in the Big PR class were observed by radial velocity (RV) and the remaining 47\% were observed by transit whereas in the small PR class, 76\% of planets were observed by transit, 14\% by RV and the remaining 10\% were observed either by transit timing variations or orbital brightness modulation.  We note these observation methods as each method has its own associated observational biases.  Most notably, the majority of the small PR planets in our sample were observed by transit which is biased against long period orbits however, the presence of additional planets could be detected by transit timing variations. On the other hand, since most of the planets in the Big PR group were observed via RV, it is possible these observations are missing additional small planets that could change the trends we describe here.  

The AMD calculations also depend on the planet inclination and eccentricity.  Because the planet inclinations are difficult to measure and hence largely absent, we set the inclination equal to zero for all planets thus making our AMD calculations only a function of the planet mass, semi-major axis, and eccentricity.  There are eccentricity measurements for 55 of the planets and so we only calculate the AMD for the planets where an eccentricity measurement is available.  This results in 39 data points for the AMD Big PR sample and 16 data points for the AMD Small PR sample.

Because 53\% of the Big PR planets were observed by radial velocity, we only have $m$sin$i$ measurements for some of the planets.  However, $m$sin$i$ is a lower limit on the mass of the planet and thus the real AMD of the big PR values should be even larger than calculated here.  On the other hand, only 10\% of the small PR planets were observed with radial velocity and have $m$sin$i$ measurements.  When calculating the adjacent planet mass ratios, we assume all planets in a system lie within the same orbital plane since we only have $m\mathrm{sin}i$ measurements for planets in some systems.

\subsection{Results}
\begin{figure*}
\includegraphics[width=0.5\textwidth]{ 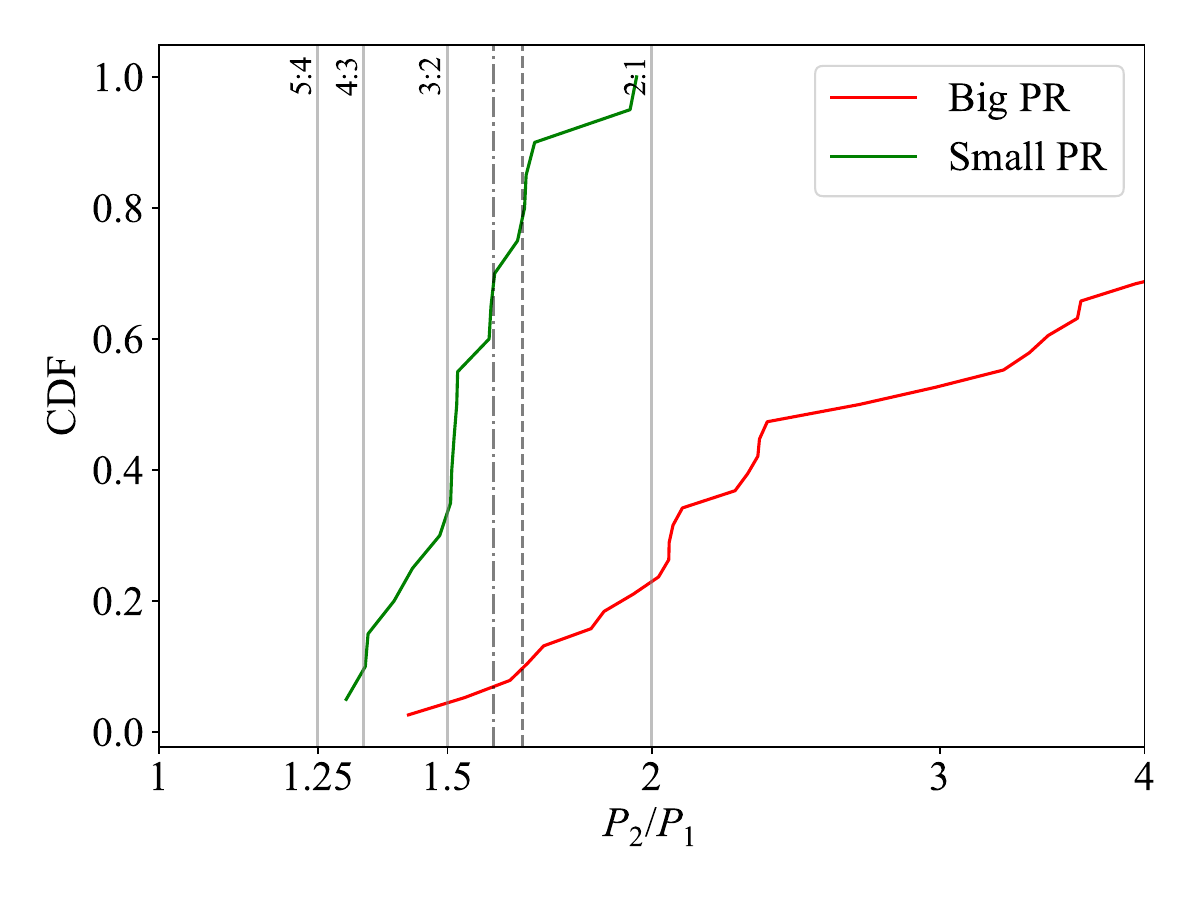}
\includegraphics[width=0.5\textwidth]{ 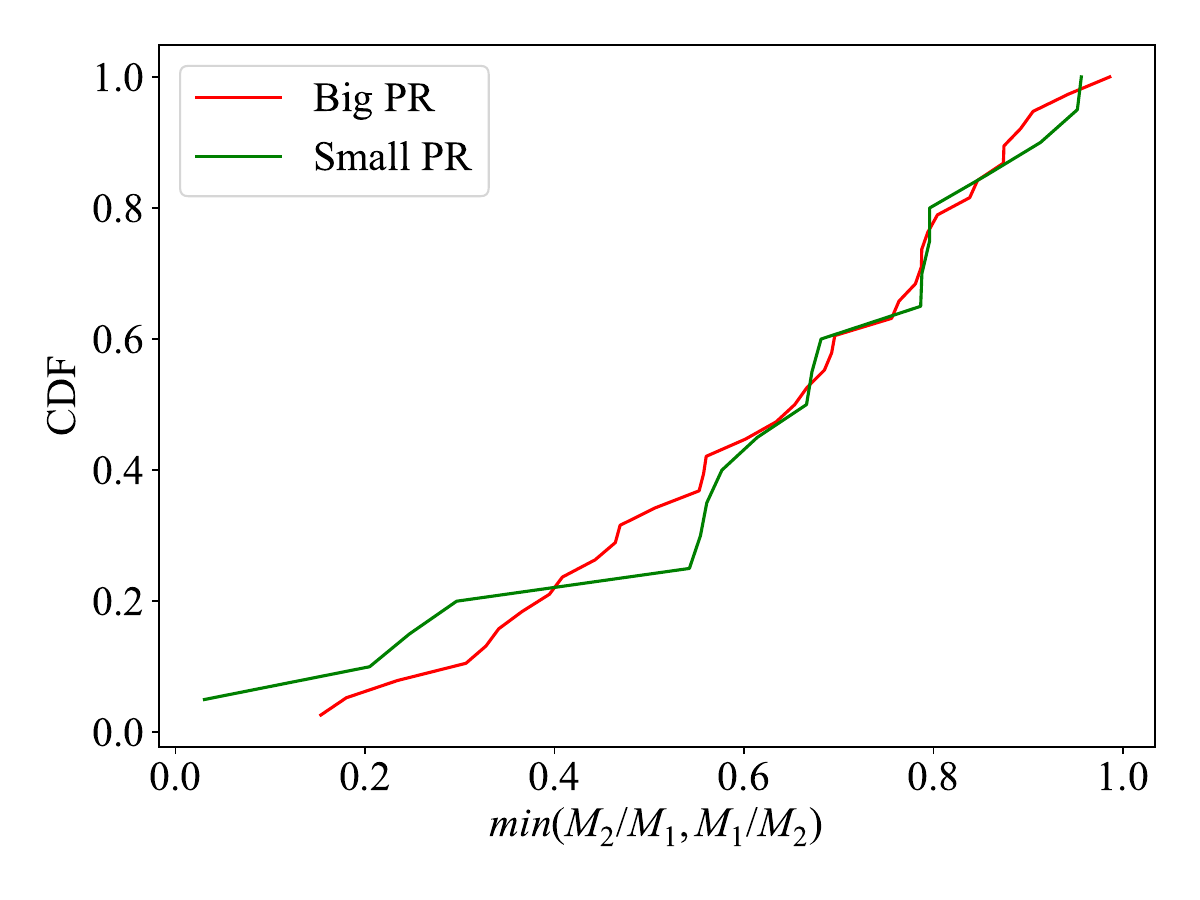}
\includegraphics[width=0.5\textwidth]{ 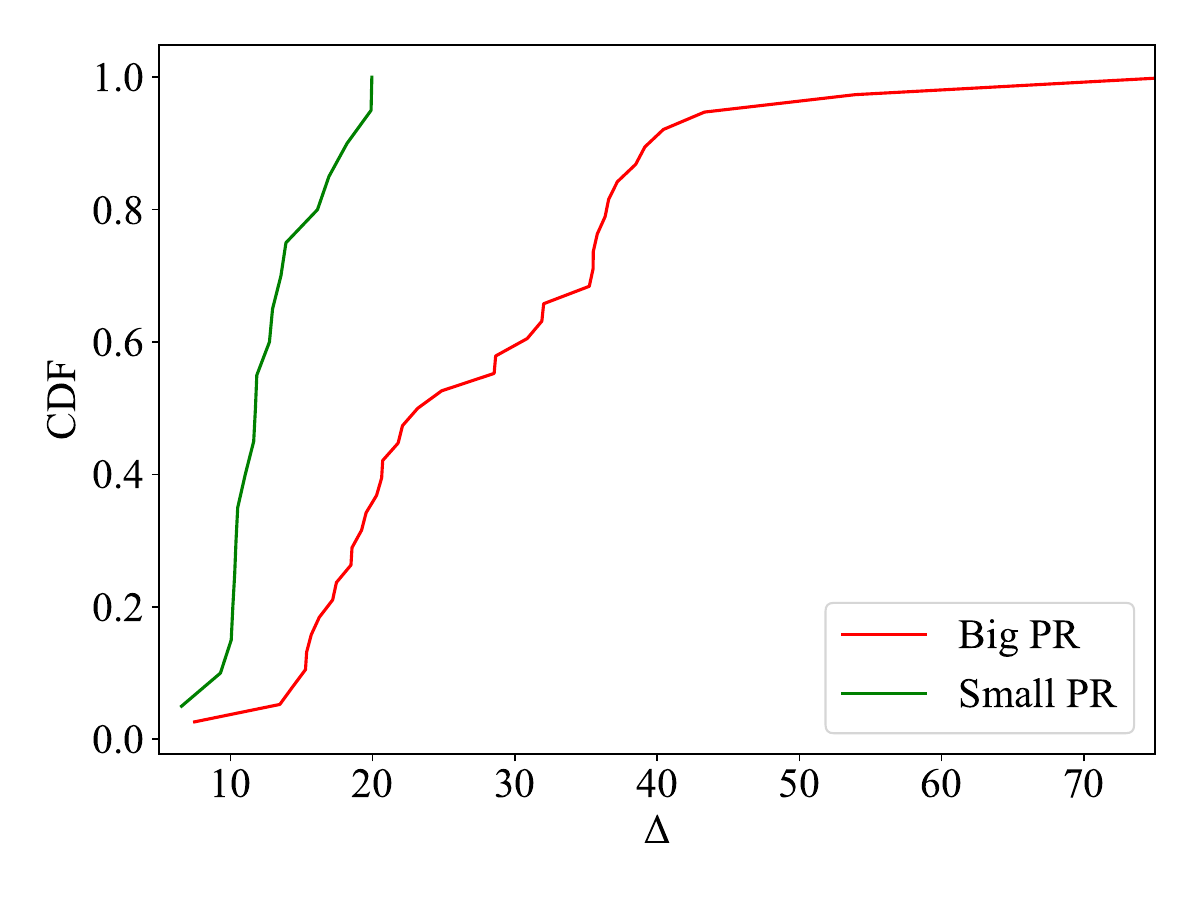}
\includegraphics[width=0.5\textwidth]{ 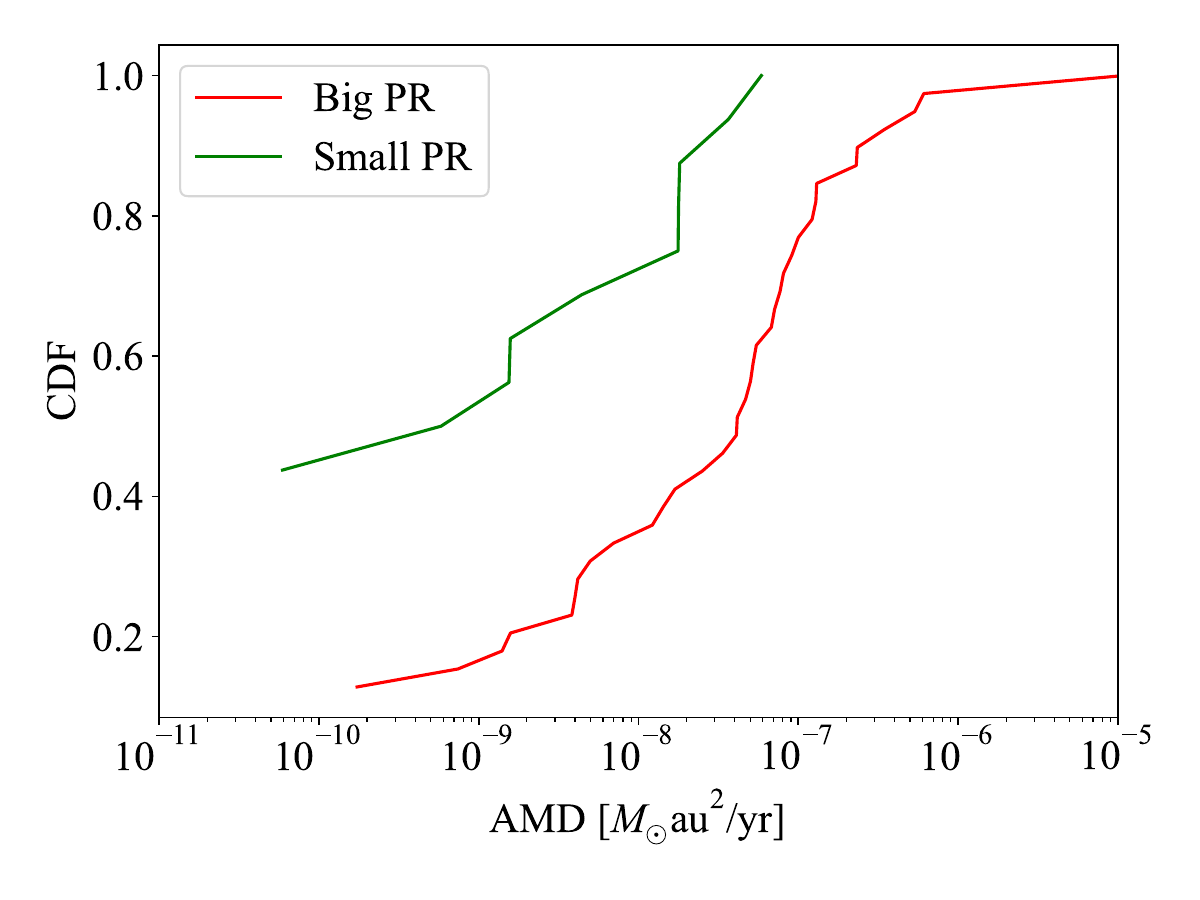}
    \caption{Observational data for the two groups of systems. Top left: CDF of period ratios for adjacent planets. The strongest first order MMRs are marked with vertical lines and the 5:3 MMR is marked with a dashed line.  Bottom left: CDF of orbital spacing between adjacent planets in units of mutual hill radii.  Top right:  CDF of mass ratio for adjacent planets.  Bottom right:  CDF of AMD for each planet.}
    \label{fig:obs_CDFs}
\end{figure*}

While the observational data sets are smaller and thus have less statistical power than our simulated data sets, we find remarkably good agreement with our expectations.  Figure \ref{fig:obs_CDFs} shows the same four stability metrics but for our two groups of observed planets.  Like simulations, we find that the distributions are statistically different for the period ratio, $\Delta$, and AMD distributions.  The period ratio, AMD, and $\Delta$ distributions have K-S test $p-$values$\ll 0.05$ between the Big PR and the Small PR groups. We note that $\sim 40 \%$ of the Small PR planets and $\sim 10 \%$ of the Big PR planets have an AMD of zero due to measured eccentricities of zero within observational error and our assumed zero inclinations. 

Similar to the simulations, we do not find a statistical difference between the observed adjacent planet mass ratios or the observed planet multiplicities.  However, we do find that on average, the Big PR planets have a \textit{slightly} smaller adjacent planet mass ratio than the Small PR planets with values of 0.619 and 0.624. Like the simulated PLM systems, the Small PR systems have a median multiplicity of three and like the simulated U systems, the Big PR systems have a median multiplicity of two. For consistency with our simulation analysis, we exclude T1 from this test since very high multiplicity systems (more than six planets) appear to be uncommon and is an indicator of no system instabilities after gas disk dissipation.

\section{Discussion and Conclusions}\label{sec:Discussion}
We have extended simulations from \cite{Childs2023} who modeled the formation of the T1 system through pebble accretion in the presence of gas.  Gas effects can stabilize compact systems by damping the eccentricity and inclination of the planets.  It also forces inward Type-I migration of the planets which commonly results in strong MMRs.  In time, the gas dissipates and  the gravitational interaction of the planets  dominates the evolution of the planetary system.  To model this, we turn off all gas effects and simulate only the gravitational interaction of the systems up to $100 \, \rm Myr$.

We classify the systems based on what their final multiplicity is and whether they experienced an instability, in the form of a planet merger or ejection, or not.  We find that 9\% of our runs are T1As, 40\% are PLMs, and 51\% experience some type of instability.  Even though the formation channel and inward migration rates for the simulations of \cite{Childs2023} were tailored to the T1 system, less than $~10\%$ of the runs were able to remain stable without gas as T1-analogs, indicating that these types of systems may be difficult to form and remain long term stable.  This may help explain why there have been no other systems similar to T1 observed.

We evaluate the structure of these systems at $3 \, \rm Myr$ in the presence of gas to assess what determines whether a system will experience an instability or not.  We found that it is the presence of three body MMRs throughout the system and the presence of a strong MMR in the innermost planet pair that largely determines whether or not a system would experience an instability after the gas disk dissipates.

Next, we looked at the structures of the systems at $100 \, \rm Myr$ to determine if there could be observational differences between the systems that experienced an instability and those that did not.  Indeed, we find significant qualitative differences between these two classes of systems. The systems that did experience an instability are more likely to lack MMRs, contain a period ratio greater than two, have smaller adjacent planet mass ratios and system multiplicities, and have larger orbital spacings ($\Delta$) and planetary AMDs, relative to the systems that did not experience an instability. This is in agreement with \cite{Pu2015} and \cite{Volk2015ApJ} who proposed that larger dynamical spacing is the result of a previous planet merger.

Lastly, we compared these findings to the observed multi-planet systems around M-dwarfs that do not host any planets larger than super-Earths to see if these same trends exist in the observations.  We broke the observed systems into two groups: Big PR and Small PR.  Big PR are planets in the systems that contain at least one period ratio greater than two and Small PR are those in the systems that do not.  We find that the same trends as we found in the simulations also persist in the observational data, indicating that there is a subset of systems that have experienced an instability after the gas disk dissipated, most likely in the form of a planet merger.  Furthermore, our results suggest that $\sim75\%$ of the observed multi-planet systems around M-dwarfs without large planets experienced some form of instability shortly after gas disk dissipation and allow us to identify the specific systems that underwent an instability later in time.  

These findings strongly agree with those of \cite{Izidoro2017, Izidoro2021} who estimated that $\sim90\%$ of observed planetary systems experienced some form of instability after gas disk dissipation.  Furthermore, \cite{Goldberg2022} also demonstrated that dynamical instabilities of resonant changes can reproduce the period ratio and mass distributions of observed planets.  While previous studies have also suggested that most planetary systems have experienced some form of planetary merger in its past, it has been largely attributed to the secular gravitational perturbations of the multi-planet system \citep{Pu2015, Volk2015ApJ}.  Here, we identified that short term instabilities that arise from the absence of gas effects can also be the culprit for scattering that results in planet mergers and/or ejections and that the presence of a period ratio larger than two is a strong indicator of such an instability.

We focus on systems that form primarily through pebble accretion and then later experience an instability.  However, systems that formed primarily through in situ core accretion may also share the same characteristics as the unstable systems described here.  While the majority of studies of in situ core accretion are around more massive solar type stars \citep{Hansen2012, Hansen2013, Dawson2016}, \cite{Gaidos2017} and \cite{Sabotta_2021} estimate the minimum mass extrasolar nebula for massive disks around M dwarfs to be $\gtrsim 5 \, M_{\oplus}$ within $0.5 \, \rm au$ during the planet formation epoch which could be sufficient for in situ planet formation.

Admittedly, we are unable to differentiate between systems that formed primarily through pebble accretion and went unstable and systems that formed primarily through late stage core accretion.  Still, in both scenarios the planets would likely have experienced late stage giant impacts which should have implications for their final properties, unlike the systems that form primarily through pebble accretion but do not experience any instabilities after the gas disk dissipates.  For example, giant impacts can produce a moon system \citep{Canup:2001aa, Canup2004}, internal heating that can result in plate tectonics and volcanism \citep{Yuan2024}, the generation of a secondary atmosphere \citep{Liggins2022}, a molten core \citep{BRIANTONKS1992326} and in turn, potentially, a magnetic field \citep{Bullard1949}.  These are all reminiscent of Earth and are thought to be beneficial to the habitability of a planet.  Thus, we suggest that systems around M-dwarfs that contain period ratios greater than two be given priority in the search for habitable worlds.


\section{Software and third party data repository citations} \label{sec:cite}

The observational data used in this study can be found in Tables \ref{tab:SMALL_PR_obs_data} and \ref{tab:BIG_PR_obs_data} and may be reproduced from the Composite Planet Data Table at \cite{NASAExoplanetArchive}.

\begin{acknowledgments}
We thank Dan Tamayo for providing comments that have improved this manuscript.  ACC and AMG acknowledge supported from the National Science Foundation (NSF) under grant No. AST-2107738. Any opinions, findings, and conclusions or recommendations expressed in this material are those of the author(s) and do not necessarily reflect the views of the NSF. RGM acknowledges support from NASA through grant 80NSSC21K0395. 
CCY acknowledges the support from NASA via the Emerging Worlds program (\#80NSSC23K0653), the Astrophysics Theory Program (grant \#80NSSC24K0133), and the Theoretical and Computational Astrophysical Networks (grant \#80NSSC21K0497).
This research was supported in part through the computational resources and staff contributions provided for the Quest high performance computing facility at Northwestern University which is jointly supported by the Office of the Provost, the Office for Research, and Northwestern University Information Technology.
\end{acknowledgments}

%

\vspace{5mm}


\software{REBOUND \citep{Rein2012}, REBOUNDx \citep{Tamayo_2021}}



 \begin{deluxetable*}{lcclcccccc}
\tablecaption{Observational data used in our study for the Small PR planets.\label{tab:SMALL_PR_obs_data}}
\tablewidth{\textwidth}
\tabletypesize{\scriptsize}
\colnumbers
\tablehead{
\colhead{Hostname} & 
\colhead{$M_{\star}/M_{\odot}$} & 
\colhead{No. Pl} & 
\colhead{Planet Name} & 
\colhead{$P_{\rm pl}$ (days)} & 
\colhead{$m_{\rm pl} / M_{\oplus}$} & 
\colhead{$m_{\rm pl} \sin i / M_{\oplus}$} & 
\colhead{$r_{\rm pl} / R_{\oplus}$} & 
\colhead{$+\sigma_1 r_{\rm pl} / R_{\oplus}$} & 
\colhead{MCMC $m_{\rm pl} / M_{\oplus}$}
}

\startdata
    K2-239 &    0.400 &        3 &     K2-239 b &      5.240 &        - &         - &   1.100 &       0.100 &       1.270 \\
    K2-239 &    0.400 &        3 &     K2-239 c &      7.775 &        - &         - &   1.000 &       0.100 &       1.011 \\
    K2-239 &    0.400 &        3 &     K2-239 d &     10.115 &        - &         - &   1.100 &       0.100 &       1.283 \\
     K2-72 &    0.270 &        4 &      K2-72 b &      5.577 &        - &         - &   1.080 &       0.110 &       1.257 \\
     K2-72 &    0.270 &        4 &      K2-72 c &     15.189 &        - &         - &   1.160 &       0.130 &       1.579 \\
     K2-72 &    0.270 &        4 &      K2-72 d &      7.760 &        - &         - &   1.010 &       0.120 &       1.076 \\
     K2-72 &    0.270 &        4 &      K2-72 e &     24.159 &        - &         - &   1.290 &       0.140 &       1.983 \\
    KOI-55 &    0.500 &        2 &     KOI-55 b &      0.240 &    0.440 &         - &   0.759 &           - &           - \\
    KOI-55 &    0.500 &        2 &     KOI-55 c &      0.343 &    0.655 &         - &   0.867 &           - &           - \\
Kepler-138 &    0.540 &        4 & Kepler-138 b &     10.313 &    0.070 &         - &   0.640 &       0.020 &           - \\
Kepler-138 &    0.540 &        4 & Kepler-138 c &     13.781 &    2.300 &         - &   1.510 &       0.040 &           - \\
Kepler-138 &    0.540 &        4 & Kepler-138 d &     23.092 &    2.100 &         - &   1.510 &       0.040 &           - \\
Kepler-138 &    0.540 &        4 & Kepler-138 e &     38.230 &    0.430 &         - &       - &           - &           - \\
Kepler-446 &    0.220 &        3 & Kepler-446 b &      1.565 &        - &         - &   1.500 &       0.250 &       2.597 \\
Kepler-446 &    0.220 &        3 & Kepler-446 c &      3.036 &        - &         - &   1.110 &       0.180 &       1.409 \\
Kepler-446 &    0.220 &        3 & Kepler-446 d &      5.149 &        - &         - &   1.350 &       0.220 &       2.443 \\
  TOI-2095 &    0.440 &        2 &   TOI-2095 b &     17.665 &    4.100 &         - &   1.250 &       0.070 &           - \\
  TOI-2095 &    0.440 &        2 &   TOI-2095 c &     28.172 &    7.400 &         - &   1.330 &       0.080 &           - \\
TRAPPIST-1 &    0.090 &        7 & TRAPPIST-1 b &      1.511 &    1.374 &         - &   1.116 &       0.014 &           - \\
TRAPPIST-1 &    0.090 &        7 & TRAPPIST-1 c &      2.422 &    1.308 &         - &   1.097 &       0.014 &           - \\
TRAPPIST-1 &    0.090 &        7 & TRAPPIST-1 d &      4.049 &    0.388 &         - &   0.788 &       0.011 &           - \\
TRAPPIST-1 &    0.090 &        7 & TRAPPIST-1 e &      6.101 &    0.692 &         - &   0.920 &       0.013 &           - \\
TRAPPIST-1 &    0.090 &        7 & TRAPPIST-1 f &      9.208 &    1.039 &         - &   1.045 &       0.013 &           - \\
TRAPPIST-1 &    0.090 &        7 & TRAPPIST-1 g &     12.352 &    1.321 &         - &   1.129 &       0.015 &           - \\
TRAPPIST-1 &    0.090 &        7 & TRAPPIST-1 h &     18.773 &    0.326 &         - &   0.755 &       0.014 &           - \\
    YZ Cet &    0.140 &        3 &     YZ Cet b &      2.021 &        - &     0.700 &       - &           - &           - \\
    YZ Cet &    0.140 &        3 &     YZ Cet c &      3.060 &        - &     1.140 &       - &           - &           - \\
    YZ Cet &    0.140 &        3 &     YZ Cet d &      4.656 &        - &     1.090 &       - &           - &           - \\
\enddata
\tablecomments{All data are from \citep{NASAExoplanetArchive}.  Column 1 lists the hostname of the system.  Column 2 lists the stellar mass of the system.  Column 3 lists the number of planets in the system.  Column 4 lists each planet name in a give system. Column 5 lists the orbital period of the planet in days.  Column 6 lists the observed mass of the planet.  Column 7 lists the observed mass*sin$i$ of the planet.  Column 8 lists the observed planet radius and column 9 is the upper error on this measurement.  Colum 10 is the approximate MCMC mass from FORECASTER \citep{Chen2017}.}
\end{deluxetable*}

 \footnotesize
\renewcommand{\arraystretch}{0.9}

\begin{deluxetable*}{lcclcccccc}
\tablecaption{Observational data used in our study for the Big PR planets.\label{tab:BIG_PR_obs_data}}
\tablewidth{\textwidth}
\tabletypesize{\scriptsize}
\colnumbers
\tablehead{
\colhead{Hostname} & 
\colhead{$M_{\star}/M_{\odot}$} & 
\colhead{No. Pl} & 
\colhead{Planet Name} & 
\colhead{$P_{\rm pl}$ (days)} & 
\colhead{$m_{\rm pl} / M_{\oplus}$} & 
\colhead{$m_{\rm pl} \sin i / M_{\oplus}$} & 
\colhead{$r_{\rm pl} / R_{\oplus}$} & 
\colhead{$+\sigma_1 r_{\rm pl} / R_{\oplus}$} & 
\colhead{MCMC $m_{\rm pl} / M_{\oplus}$}
}
\startdata
       G 264-012 &    0.300 &        2 &        G 264-012 b &      2.305 &        - &     2.500 &       - &           - &           - \\
       G 264-012 &    0.300 &        2 &        G 264-012 c &      8.052 &        - &     3.750 &       - &           - &           - \\
         GJ 1002 &    0.120 &        2 &          GJ 1002 b &     10.347 &        - &     1.080 &       - &           - &           - \\
         GJ 1002 &    0.120 &        2 &          GJ 1002 c &     21.202 &        - &     1.360 &       - &           - &           - \\
         GJ 1061 &    0.120 &        3 &          GJ 1061 b &      3.204 &        - &     1.370 &       - &           - &           - \\
         GJ 1061 &    0.120 &        3 &          GJ 1061 c &      6.689 &        - &     1.740 &       - &           - &           - \\
         GJ 1061 &    0.120 &        3 &          GJ 1061 d &     13.031 &        - &     1.640 &       - &           - &           - \\
         GJ 1132 &    0.190 &        2 &          GJ 1132 b &      1.629 &    1.837 &     1.837 &   1.192 &       0.038 &           - \\
         GJ 1132 &    0.180 &        2 &          GJ 1132 c &      8.929 &        - &     2.640 &       - &           - &           - \\
          GJ 180 &    0.430 &        3 &           GJ 180 b &     17.133 &        - &     6.490 &       - &           - &           - \\
          GJ 180 &    0.430 &        3 &           GJ 180 c &     24.329 &        - &     6.400 &       - &           - &           - \\
          GJ 180 &    0.430 &        3 &           GJ 180 d &    106.300 &        - &     7.560 &       - &           - &           - \\
          GJ 273 &    0.290 &        2 &           GJ 273 b &     18.650 &        - &     2.890 &       - &           - &           - \\
          GJ 273 &    0.290 &        2 &           GJ 273 c &      4.723 &        - &     1.180 &       - &           - &           - \\
         GJ 3323 &    0.160 &        2 &          GJ 3323 b &      5.364 &        - &     2.020 &       - &           - &           - \\
         GJ 3323 &    0.160 &        2 &          GJ 3323 c &     40.540 &        - &     2.310 &       - &           - &           - \\
          GJ 357 &    0.340 &        3 &           GJ 357 b &      3.931 &        - &         - &   1.200 &       0.060 &       1.595 \\
          GJ 357 &    0.340 &        3 &           GJ 357 c &      9.125 &        - &     3.400 &       - &           - &           - \\
          GJ 357 &    0.340 &        3 &           GJ 357 d &     55.661 &        - &     6.100 &       - &           - &           - \\
          GJ 367 &    0.460 &        3 &           GJ 367 b &      0.322 &    0.633 &         - &   0.699 &       0.024 &           - \\
          GJ 367 &    0.460 &        3 &           GJ 367 c &     11.530 &        - &     4.130 &       - &           - &           - \\
          GJ 367 &    0.460 &        3 &           GJ 367 d &     34.369 &        - &     6.030 &       - &           - &           - \\
         GJ 3929 &    0.310 &        2 &          GJ 3929 b &      2.616 &    1.750 &         - &   1.090 &       0.040 &           - \\
         GJ 3929 &    0.310 &        2 &          GJ 3929 c &     15.040 &        - &     5.710 &       - &           - &           - \\
         GJ 3998 &    0.500 &        2 &          GJ 3998 b &      2.650 &        - &     2.470 &       - &           - &           - \\
         GJ 3998 &    0.500 &        2 &          GJ 3998 c &     13.740 &        - &     6.260 &       - &           - &           - \\
          GJ 682 &    0.270 &        2 &           GJ 682 b &     17.478 &        - &     4.400 &       - &           - &           - \\
          GJ 682 &    0.270 &        2 &           GJ 682 c &     57.320 &        - &     8.700 &       - &           - &           - \\
          GJ 806 &    0.410 &        2 &           GJ 806 b &      0.926 &    1.900 &         - &   1.331 &       0.023 &           - \\
          GJ 806 &    0.410 &        2 &           GJ 806 c &      6.641 &        - &     5.800 &       - &           - &           - \\
          GJ 887 &    0.490 &        2 &           GJ 887 b &      9.262 &        - &     4.200 &       - &           - &           - \\
          GJ 887 &    0.490 &        2 &           GJ 887 c &     21.789 &        - &     7.600 &       - &           - &           - \\
       HD 260655 &    0.440 &        2 &        HD 260655 b &      2.770 &    2.140 &         - &   1.240 &       0.023 &           - \\
       HD 260655 &    0.440 &        2 &        HD 260655 c &      5.706 &    3.090 &         - &   1.533 &       0.051 &           - \\
          K2-240 &    0.580 &        2 &           K2-240 b &      6.034 &        - &         - &   2.000 &       0.200 &       5.203 \\
          K2-240 &    0.580 &        2 &           K2-240 c &     20.523 &        - &         - &   1.800 &       0.300 &       4.185 \\
          K2-316 &    0.410 &        2 &           K2-316 b &      1.133 &        - &         - &   1.330 &       0.100 &       2.270 \\
          K2-316 &    0.410 &        2 &           K2-316 c &      5.260 &        - &         - &   1.830 &       0.130 &       4.890 \\
           K2-83 &    0.480 &        2 &            K2-83 b &      2.747 &        - &         - &   1.244 &       0.098 &       2.046 \\
           K2-83 &    0.480 &        2 &            K2-83 c &      9.998 &        - &         - &   1.483 &       0.099 &       3.401 \\
     Kepler-1649 &    0.200 &        2 &      Kepler-1649 b &      8.689 &        - &         - &   1.017 &       0.051 &       1.076 \\
     Kepler-1649 &    0.200 &        2 &      Kepler-1649 c &     19.535 &        - &         - &   1.060 &       0.150 &       1.206 \\
      Kepler-186 &    0.480 &        5 &       Kepler-186 b &      3.887 &        - &         - &   1.070 &       0.120 &       1.351 \\
      Kepler-186 &    0.480 &        5 &       Kepler-186 c &      7.267 &        - &         - &   1.250 &       0.140 &       1.788 \\
      Kepler-186 &    0.480 &        5 &       Kepler-186 d &     13.343 &        - &         - &   1.400 &       0.160 &       2.736 \\
      Kepler-186 &    0.480 &        5 &       Kepler-186 e &     22.408 &        - &         - &   1.270 &       0.150 &       2.293 \\
      Kepler-186 &    0.540 &        5 &       Kepler-186 f &    129.944 &        - &         - &   1.170 &       0.080 &       1.751 \\
       Kepler-42 &    0.130 &        3 &        Kepler-42 b &      1.214 &        - &         - &   0.780 &       0.220 &       0.505 \\
       Kepler-42 &    0.130 &        3 &        Kepler-42 c &      0.453 &        - &         - &   0.730 &       0.200 &       0.397 \\
       Kepler-42 &    0.130 &        3 &        Kepler-42 d &      1.865 &        - &         - &   0.570 &       0.180 &       0.118 \\
         L 98-59 &    0.270 &        4 &          L 98-59 b &      2.253 &    0.400 &         - &   0.850 &       0.061 &           - \\
         L 98-59 &    0.270 &        4 &          L 98-59 c &      3.691 &    2.220 &         - &   1.385 &       0.095 &           - \\
         L 98-59 &    0.270 &        4 &          L 98-59 d &      7.451 &    1.940 &         - &   1.521 &       0.119 &           - \\
         L 98-59 &    0.270 &        4 &          L 98-59 e &     12.796 &        - &     3.060 &       - &           - &           - \\
        LHS 1140 &    0.180 &        2 &         LHS 1140 b &     24.737 &    5.600 &         - &   1.730 &       0.025 &           - \\
        LHS 1140 &    0.180 &        2 &         LHS 1140 c &      3.778 &    1.910 &         - &   1.272 &       0.026 &           - \\
        TOI-2096 &    0.230 &        2 &         TOI-2096 b &      3.119 &        - &         - &   1.243 &       0.077 &       1.883 \\
        TOI-2096 &    0.230 &        2 &         TOI-2096 c &      6.388 &        - &         - &   1.914 &       0.095 &       5.149 \\
Teegarden's Star &    0.100 &        3 & Teegarden's Star b &      4.906 &        - &     1.160 &       - &           - &           - \\
Teegarden's Star &    0.100 &        3 & Teegarden's Star c &     11.416 &        - &     1.050 &       - &           - &           - \\
Teegarden's Star &    0.100 &        3 & Teegarden's Star d &     26.130 &        - &     0.820 &       - &           - &           - \\
       Wolf 1061 &    0.290 &        3 &        Wolf 1061 b &      4.887 &        - &     1.910 &       - &           - &           - \\
       Wolf 1061 &    0.290 &        3 &        Wolf 1061 c &     17.872 &        - &     3.410 &       - &           - &           - \\
       Wolf 1061 &    0.290 &        3 &        Wolf 1061 d &    217.210 &        - &     7.700 &       - &           - &           - \\
\enddata
\tablecomments{All data are from \citep{NASAExoplanetArchive}.  Column 1 lists the hostname of the system.  Column 2 lists the stellar mass of the system.  Column 3 lists the number of planets in the system.  Column 4 lists each planet name in a give system. Column 5 lists the orbital period of the planet in days.  Column 6 lists the observed mass of the planet.  Column 7 lists the observed mass*sin$i$ of the planet.  Column 8 lists the observed planet radius and column 9 is the upper error on this measurement.  Colum 10 is the approximate MCMC mass from FORECASTER \citep{Chen2017}.}
\end{deluxetable*}


\bibliography{main}{}

\begin{thebibliography}{}
\expandafter\ifx\csname natexlab\endcsname\relax\def\natexlab#1{#1}\fi
\providecommand{\url}[1]{\href{#1}{#1}}
\providecommand{\dodoi}[1]{doi:~\href{http://doi.org/#1}{\nolinkurl{#1}}}
\providecommand{\doeprint}[1]{\href{http://ascl.net/#1}{\nolinkurl{http://ascl.net/#1}}}
\providecommand{\doarXiv}[1]{\href{https://arxiv.org/abs/#1}{\nolinkurl{https://arxiv.org/abs/#1}}}

\bibitem[{{Agol} {et~al.}(2021){Agol}, {Dorn}, {Grimm}, {Turbet}, {Ducrot}, {Delrez}, {Gillon}, {Demory}, {Burdanov}, {Barkaoui}, {Benkhaldoun}, {Bolmont}, {Burgasser}, {Carey}, {de Wit}, {Fabrycky}, {Foreman-Mackey}, {Haldemann}, {Hernandez}, {Ingalls}, {Jehin}, {Langford}, {Leconte}, {Lederer}, {Luger}, {Malhotra}, {Meadows}, {Morris}, {Pozuelos}, {Queloz}, {Raymond}, {Selsis}, {Sestovic}, {Triaud}, \& {Van Grootel}}]{Agol2021}
{Agol}, E., {Dorn}, C., {Grimm}, S.~L., {et~al.} 2021, \psj, 2, 1, \dodoi{10.3847/PSJ/abd022}

\bibitem[{{Brian Tonks} \& {Jay Melosh}(1992)}]{BRIANTONKS1992326}
{Brian Tonks}, W., \& {Jay Melosh}, H. 1992, Icarus, 100, 326

\bibitem[{{Bullard}(1949)}]{Bullard1949}
{Bullard}, E.~C. 1949, Proceedings of the Royal Society of London Series A, 197, 433, \dodoi{10.1098/rspa.1949.0074}

\bibitem[{{Burgasser} \& {Mamajek}(2017)}]{Burgasser2017}
{Burgasser}, A.~J., \& {Mamajek}, E.~E. 2017, \apj, 845, 110, \dodoi{10.3847/1538-4357/aa7fea}

\bibitem[{{Canup}(2004)}]{Canup2004}
{Canup}, R.~M. 2004, \icarus, 168, 433, \dodoi{10.1016/j.icarus.2003.09.028}

\bibitem[{Canup \& Asphaug(2001)}]{Canup:2001aa}
Canup, R.~M., \& Asphaug, E. 2001, Nature, 412, 708

\bibitem[{{Chen} \& {Kipping}(2017)}]{Chen2017}
{Chen}, J., \& {Kipping}, D. 2017, \apj, 834, 17, \dodoi{10.3847/1538-4357/834/1/17}

\bibitem[{{Childs} {et~al.}(2022){Childs}, {Martin}, \& {Livio}}]{Childs2022mdwarf}
{Childs}, A.~C., {Martin}, R.~G., \& {Livio}, M. 2022, \apjl, 937, L41, \dodoi{10.3847/2041-8213/ac9052}

\bibitem[{{Childs} {et~al.}(2023){Childs}, {Shakespeare}, {Rice}, {Yang}, \& {Steffen}}]{Childs2023}
{Childs}, A.~C., {Shakespeare}, C., {Rice}, D.~R., {Yang}, C.-C., \& {Steffen}, J.~H. 2023, \mnras, 524, 3749, \dodoi{10.1093/mnras/stad2110}

\bibitem[{{Childs} \& {Steffen}(2022)}]{Childs_Steffen_2022}
{Childs}, A.~C., \& {Steffen}, J.~H. 2022, \mnras, 511, 1848, \dodoi{10.1093/mnras/stac158}

\bibitem[{{Clement} {et~al.}(2024){Clement}, {Quintana}, \& {Stevenson}}]{Clement2024}
{Clement}, M.~S., {Quintana}, E.~V., \& {Stevenson}, K.~B. 2024, arXiv e-prints, arXiv:2411.02529, \dodoi{10.48550/arXiv.2411.02529}

\bibitem[{{Coleman} {et~al.}(2019){Coleman}, {Leleu}, {Alibert}, \& {Benz}}]{Coleman2019}
{Coleman}, G.~A.~L., {Leleu}, A., {Alibert}, Y., \& {Benz}, W. 2019, \aap, 631, A7, \dodoi{10.1051/0004-6361/201935922}

\bibitem[{{Cresswell} \& {Nelson}(2006)}]{Cresswell_2006}
{Cresswell}, P., \& {Nelson}, R.~P. 2006, \aap, 450, 833, \dodoi{10.1051/0004-6361:20054551}

\bibitem[{{Cresswell} \& {Nelson}(2008)}]{Cresswell2008}
---. 2008, \aap, 482, 677, \dodoi{10.1051/0004-6361:20079178}

\bibitem[{{Dawson} {et~al.}(2016){Dawson}, {Lee}, \& {Chiang}}]{Dawson2016}
{Dawson}, R.~I., {Lee}, E.~J., \& {Chiang}, E. 2016, \apj, 822, 54, \dodoi{10.3847/0004-637X/822/1/54}

\bibitem[{{Fang} \& {Margot}(2013)}]{Fang2013F}
{Fang}, J., \& {Margot}, J.-L. 2013, \apj, 767, 115, \dodoi{10.1088/0004-637X/767/2/115}

\bibitem[{{Fischer} \& {Valenti}(2005)}]{Fischer2005}
{Fischer}, D.~A., \& {Valenti}, J. 2005, \apj, 622, 1102, \dodoi{10.1086/428383}

\bibitem[{{Gaidos}(2017)}]{Gaidos2017}
{Gaidos}, E. 2017, \mnras, 470, L1, \dodoi{10.1093/mnrasl/slx063}

\bibitem[{Gallardo {et~al.}(2016)Gallardo, Coito, \& Badano}]{Gallardo2016}
Gallardo, T., Coito, L., \& Badano, L. 2016, Icarus, 274, 83

\bibitem[{{Ghezzi} {et~al.}(2010){Ghezzi}, {Cunha}, {Schuler}, \& {Smith}}]{Ghezzi2010}
{Ghezzi}, L., {Cunha}, K., {Schuler}, S.~C., \& {Smith}, V.~V. 2010, \apj, 725, 721, \dodoi{10.1088/0004-637X/725/1/721}

\bibitem[{{Ghezzi} {et~al.}(2018){Ghezzi}, {Montet}, \& {Johnson}}]{Ghezzi2018}
{Ghezzi}, L., {Montet}, B.~T., \& {Johnson}, J.~A. 2018, \apj, 860, 109, \dodoi{10.3847/1538-4357/aac37c}

\bibitem[{{Gillon} {et~al.}(2016){Gillon}, {Jehin}, {Lederer}, {Delrez}, {de Wit}, {Burdanov}, {Van Grootel}, {Burgasser}, {Triaud}, {Opitom}, {Demory}, {Sahu}, {Bardalez Gagliuffi}, {Magain}, \& {Queloz}}]{Gillon2016}
{Gillon}, M., {Jehin}, E., {Lederer}, S.~M., {et~al.} 2016, \nat, 533, 221, \dodoi{10.1038/nature17448}

\bibitem[{Gillon {et~al.}(2017)Gillon, Triaud, Demory, Jehin, Agol, Deck, Lederer, de~Wit, Burdanov, Ingalls, Bolmont, Leconte, Raymond, Selsis, Turbet, Barkaoui, Burgasser, Burleigh, Carey, Chaushev, Copperwheat, Delrez, Fernandes, Holdsworth, Kotze, Van~Grootel, Almleaky, Benkhaldoun, Magain, \& Queloz}]{Gillon2017}
Gillon, M., Triaud, A. H. M.~J., Demory, B.-O., {et~al.} 2017, Nature, 542, 456

\bibitem[{{Goldberg} \& {Batygin}(2022)}]{Goldberg2022}
{Goldberg}, M., \& {Batygin}, K. 2022, \aj, 163, 201, \dodoi{10.3847/1538-3881/ac5961}

\bibitem[{{Goldreich} \& {Schlichting}(2014)}]{Goldreich2014}
{Goldreich}, P., \& {Schlichting}, H.~E. 2014, \aj, 147, 32, \dodoi{10.1088/0004-6256/147/2/32}

\bibitem[{{Gonzalez}(1997)}]{Gonzalez1997}
{Gonzalez}, G. 1997, \mnras, 285, 403, \dodoi{10.1093/mnras/285.2.403}

\bibitem[{{Greene} {et~al.}(2023){Greene}, {Bell}, {Ducrot}, {Dyrek}, {Lagage}, \& {Fortney}}]{Greene2023}
{Greene}, T.~P., {Bell}, T.~J., {Ducrot}, E., {et~al.} 2023, \nat, 618, 39, \dodoi{10.1038/s41586-023-05951-7}

\bibitem[{{Grimm} {et~al.}(2018){Grimm}, {Demory}, {Gillon}, {Dorn}, {Agol}, {Burdanov}, {Delrez}, {Sestovic}, {Triaud}, {Turbet}, {Bolmont}, {Caldas}, {de Wit}, {Jehin}, {Leconte}, {Raymond}, {Van Grootel}, {Burgasser}, {Carey}, {Fabrycky}, {Heng}, {Hernandez}, {Ingalls}, {Lederer}, {Selsis}, \& {Queloz}}]{Grimm2018}
{Grimm}, S.~L., {Demory}, B.-O., {Gillon}, M., {et~al.} 2018, \aap, 613, A68, \dodoi{10.1051/0004-6361/201732233}

\bibitem[{{Hansen} \& {Murray}(2012)}]{Hansen2012}
{Hansen}, B. M.~S., \& {Murray}, N. 2012, \apj, 751, 158, \dodoi{10.1088/0004-637X/751/2/158}

\bibitem[{{Hansen} \& {Murray}(2013)}]{Hansen2013}
---. 2013, \apj, 775, 53, \dodoi{10.1088/0004-637X/775/1/53}

\bibitem[{{Hartmann} {et~al.}(1998){Hartmann}, {Calvet}, {Gullbring}, \& {D'Alessio}}]{Hartmann1998}
{Hartmann}, L., {Calvet}, N., {Gullbring}, E., \& {D'Alessio}, P. 1998, \apj, 495, 385, \dodoi{10.1086/305277}

\bibitem[{Hoshino \& Kokubo(2022)}]{Hoshino2023}
Hoshino, H., \& Kokubo, E. 2022, Monthly Notices of the Royal Astronomical Society, 519, 2838

\bibitem[{Huang \& Ormel(2022)}]{Huang2022}
Huang, S., \& Ormel, C.~W. 2022, Monthly Notices of the Royal Astronomical Society, 511, 3814

\bibitem[{Ida \& Lin(2010)}]{ida2010}
Ida, S., \& Lin, D. 2010, The Astrophysical Journal, 719, 810

\bibitem[{{Izidoro} {et~al.}(2021){Izidoro}, {Bitsch}, {Raymond}, {Johansen}, {Morbidelli}, {Lambrechts}, \& {Jacobson}}]{Izidoro2021}
{Izidoro}, A., {Bitsch}, B., {Raymond}, S.~N., {et~al.} 2021, \aap, 650, A152, \dodoi{10.1051/0004-6361/201935336}

\bibitem[{{Izidoro} {et~al.}(2017){Izidoro}, {Ogihara}, {Raymond}, {Morbidelli}, {Pierens}, {Bitsch}, {Cossou}, \& {Hersant}}]{Izidoro2017}
{Izidoro}, A., {Ogihara}, M., {Raymond}, S.~N., {et~al.} 2017, \mnras, 470, 1750, \dodoi{10.1093/mnras/stx1232}

\bibitem[{{Johansen} {et~al.}(2014){Johansen}, {Blum}, {Tanaka}, {Ormel}, {Bizzarro}, \& {Rickman}}]{Johansen2014}
{Johansen}, A., {Blum}, J., {Tanaka}, H., {et~al.} 2014, in Protostars and Planets VI, ed. H.~{Beuther}, R.~S. {Klessen}, C.~P. {Dullemond}, \& T.~{Henning}, 547--570, \dodoi{10.2458/azu_uapress_9780816531240-ch024}

\bibitem[{Johansen \& Lambrechts(2017)}]{Johansen2017}
Johansen, A., \& Lambrechts, M. 2017, Annual Review of Earth and Planetary Sciences, 45, 359

\bibitem[{Liggins {et~al.}(2022)Liggins, Jordan, Rimmer, \& Shorttle}]{Liggins2022}
Liggins, P., Jordan, S., Rimmer, P.~B., \& Shorttle, O. 2022, Journal of Geophysical Research: Planets, 127, e2021JE007123

\bibitem[{{Lim} {et~al.}(2023){Lim}, {Benneke}, {Doyon}, {MacDonald}, {Piaulet}, {Artigau}, {Coulombe}, {Radica}, {L'Heureux}, {Albert}, {Rackham}, {de Wit}, {Salhi}, {Roy}, {Flagg}, {Fournier-Tondreau}, {Taylor}, {Cook}, {Lafreni{\`e}re}, {Cowan}, {Kaltenegger}, {Rowe}, {Espinoza}, {Dang}, \& {Darveau-Bernier}}]{Lim2023}
{Lim}, O., {Benneke}, B., {Doyon}, R., {et~al.} 2023, \apjl, 955, L22, \dodoi{10.3847/2041-8213/acf7c4}

\bibitem[{Lincowski {et~al.}(2018)Lincowski, Meadows, Crisp, Robinson, Luger, Lustig-Yaeger, \& Arney}]{Lincowski_2018}
Lincowski, A.~P., Meadows, V.~S., Crisp, D., {et~al.} 2018, The Astrophysical Journal, 867, 76

\bibitem[{{Luger} {et~al.}(2017){Luger}, {Sestovic}, {Kruse}, {Grimm}, {Demory}, {Agol}, {Bolmont}, {Fabrycky}, {Fernandes}, {Van Grootel}, {Burgasser}, {Gillon}, {Ingalls}, {Jehin}, {Raymond}, {Selsis}, {Triaud}, {Barclay}, {Barentsen}, {Howell}, {Delrez}, {de Wit}, {Foreman-Mackey}, {Holdsworth}, {Leconte}, {Lederer}, {Turbet}, {Almleaky}, {Benkhaldoun}, {Magain}, {Morris}, {Heng}, \& {Queloz}}]{Luger2017}
{Luger}, R., {Sestovic}, M., {Kruse}, E., {et~al.} 2017, Nature Astronomy, 1, 0129, \dodoi{10.1038/s41550-017-0129}

\bibitem[{{McNeil} {et~al.}(2005){McNeil}, {Duncan}, \& {Levison}}]{McNeil_2005}
{McNeil}, D., {Duncan}, M., \& {Levison}, H.~F. 2005, \aj, 130, 2884, \dodoi{10.1086/497687}

\bibitem[{{Mulders} {et~al.}(2021){Mulders}, {Dr{\.z}kowska}, {van der Marel}, {Ciesla}, \& {Pascucci}}]{Mulders2021}
{Mulders}, G.~D., {Dr{\.z}kowska}, J., {van der Marel}, N., {Ciesla}, F.~J., \& {Pascucci}, I. 2021, \apjl, 920, L1, \dodoi{10.3847/2041-8213/ac2947}

\bibitem[{{NASA Exoplanet Archive}(2019)}]{NASAExoplanetArchive}
{NASA Exoplanet Archive}. 2019, Composite Planet Data Table,  IPAC, \dodoi{10.26133/NEA2}

\bibitem[{{Nesvorn{\'y}} \& {Morbidelli}(1998)}]{Nesvorny1998}
{Nesvorn{\'y}}, D., \& {Morbidelli}, A. 1998, \aj, 116, 3029, \dodoi{10.1086/300632}

\bibitem[{Ogihara \& Ida(2009)}]{Ogihara_2009}
Ogihara, M., \& Ida, S. 2009, The Astrophysical Journal, 699, 824

\bibitem[{{Ormel} {et~al.}(2017){Ormel}, {Liu}, \& {Schoonenberg}}]{Ormel2017}
{Ormel}, C.~W., {Liu}, B., \& {Schoonenberg}, D. 2017, \aap, 604, A1, \dodoi{10.1051/0004-6361/201730826}

\bibitem[{{Petit} {et~al.}(2020){Petit}, {Pichierri}, {Davies}, \& {Johansen}}]{Petit2020}
{Petit}, A.~C., {Pichierri}, G., {Davies}, M.~B., \& {Johansen}, A. 2020, \aap, 641, A176, \dodoi{10.1051/0004-6361/202038764}

\bibitem[{Petrovich {et~al.}(2014)Petrovich, Tremaine, \& Rafikov}]{Petrovich_2014}
Petrovich, C., Tremaine, S., \& Rafikov, R. 2014, The Astrophysical Journal, 786, 101

\bibitem[{{Pichierri} \& {Morbidelli}(2020)}]{Pichierri2020}
{Pichierri}, G., \& {Morbidelli}, A. 2020, \mnras, 494, 4950, \dodoi{10.1093/mnras/staa1102}

\bibitem[{{Pichierri} {et~al.}(2024){Pichierri}, {Morbidelli}, {Batygin}, \& {Brasser}}]{Pichierri2024}
{Pichierri}, G., {Morbidelli}, A., {Batygin}, K., \& {Brasser}, R. 2024, Nature Astronomy, 8, 1408, \dodoi{10.1038/s41550-024-02342-4}

\bibitem[{{Pu} \& {Wu}(2015)}]{Pu2015}
{Pu}, B., \& {Wu}, Y. 2015, \apj, 807, 44, \dodoi{10.1088/0004-637X/807/1/44}

\bibitem[{{Quarles} {et~al.}(2017){Quarles}, {Quintana}, {Lopez}, {Schlieder}, \& {Barclay}}]{Quarles2017}
{Quarles}, B., {Quintana}, E.~V., {Lopez}, E., {Schlieder}, J.~E., \& {Barclay}, T. 2017, \apjl, 842, L5, \dodoi{10.3847/2041-8213/aa74bf}

\bibitem[{{Quillen}(2011)}]{Quillen2011}
{Quillen}, A.~C. 2011, \mnras, 418, 1043, \dodoi{10.1111/j.1365-2966.2011.19555.x}

\bibitem[{{Quintana} {et~al.}(2016){Quintana}, {Barclay}, {Borucki}, {Rowe}, \& {Chambers}}]{Quintana2016}
{Quintana}, E.~V., {Barclay}, T., {Borucki}, W.~J., {Rowe}, J.~F., \& {Chambers}, J.~E. 2016, \apj, 821, 126, \dodoi{10.3847/0004-637X/821/2/126}

\bibitem[{{Rath} {et~al.}(2022){Rath}, {Hadden}, \& {Lithwick}}]{Rath2022}
{Rath}, J., {Hadden}, S., \& {Lithwick}, Y. 2022, \apj, 932, 61, \dodoi{10.3847/1538-4357/ac5f57}

\bibitem[{{Raymond} {et~al.}(2007){Raymond}, {Scalo}, \& {Meadows}}]{Raymond2007}
{Raymond}, S.~N., {Scalo}, J., \& {Meadows}, V.~S. 2007, \apj, 669, 606, \dodoi{10.1086/521587}

\bibitem[{{Raymond} {et~al.}(2022){Raymond}, {Izidoro}, {Bolmont}, {Dorn}, {Selsis}, {Turbet}, {Agol}, {Barth}, {Carone}, {Dasgupta}, {Gillon}, \& {Grimm}}]{Raymond2022}
{Raymond}, S.~N., {Izidoro}, A., {Bolmont}, E., {et~al.} 2022, Nature Astronomy, 6, 80, \dodoi{10.1038/s41550-021-01518-6}

\bibitem[{{Rein} \& {Liu}(2012)}]{Rein2012}
{Rein}, H., \& {Liu}, S.~F. 2012, \aap, 537, A128, \dodoi{10.1051/0004-6361/201118085}

\bibitem[{Rice \& Steffen(2023)}]{Rice2023}
Rice, D.~R., \& Steffen, J.~H. 2023, Monthly Notices of the Royal Astronomical Society, 520, 4057

\bibitem[{{Rosenthal} {et~al.}(2019){Rosenthal}, {Jacobson-Galan}, {Nelson}, {Murray-Clay}, {Burt}, {Holden}, {Chang}, {Kaaz}, {Yant}, {Butler}, \& {Vogt}}]{Rosenthal2019}
{Rosenthal}, M.~M., {Jacobson-Galan}, W., {Nelson}, B., {et~al.} 2019, \aj, 158, 136, \dodoi{10.3847/1538-3881/ab3b02}

\bibitem[{Sabotta {et~al.}(2021)Sabotta, Schlecker, Chaturvedi, Guenther, Muñoz~Rodríguez, Muñoz~Sánchez, Caballero, Shan, Reffert, Ribas, Reiners, Hatzes, Amado, Klahr, Morales, Quirrenbach, Henning, Dreizler, Pallé, Perger, Azzaro, Jeffers, Kaminski, Kürster, Lafarga, Montes, Passegger, \& Zechmeister}]{Sabotta_2021}
Sabotta, S., Schlecker, M., Chaturvedi, P., {et~al.} 2021, Astronomy \& Astrophysics, 653, A114, \dodoi{10.1051/0004-6361/202140968}

\bibitem[{Schoonenberg {et~al.}(2019)Schoonenberg, Liu, Ormel, \& Dorn}]{Schoonenberg_2019}
Schoonenberg, D., Liu, B., Ormel, C.~W., \& Dorn, C. 2019, Astronomy \& Astrophysics, 627, A149, \dodoi{10.1051/0004-6361/201935607}

\bibitem[{{Steffen} {et~al.}(2013){Steffen}, {Fabrycky}, {Agol}, {Ford}, {Morehead}, {Cochran}, {Lissauer}, {Adams}, {Borucki}, {Bryson}, {Caldwell}, {Dupree}, {Jenkins}, {Robertson}, {Rowe}, {Seader}, {Thompson}, \& {Twicken}}]{Steffen2013}
{Steffen}, J.~H., {Fabrycky}, D.~C., {Agol}, E., {et~al.} 2013, \mnras, 428, 1077, \dodoi{10.1093/mnras/sts090}

\bibitem[{Tamayo {et~al.}(2021)Tamayo, Murray, Tremaine, \& Winn}]{Tamayo_2021}
Tamayo, D., Murray, N., Tremaine, S., \& Winn, J. 2021, The Astronomical Journal, 162, 220

\bibitem[{{Tamayo} {et~al.}(2017){Tamayo}, {Rein}, {Petrovich}, \& {Murray}}]{Tamayo2017}
{Tamayo}, D., {Rein}, H., {Petrovich}, C., \& {Murray}, N. 2017, \apjl, 840, L19, \dodoi{10.3847/2041-8213/aa70ea}

\bibitem[{{Tamayo} {et~al.}(2015){Tamayo}, {Triaud}, {Menou}, \& {Rein}}]{Tamayo2015}
{Tamayo}, D., {Triaud}, A.~H.~M.~J., {Menou}, K., \& {Rein}, H. 2015, \apj, 805, 100, \dodoi{10.1088/0004-637X/805/2/100}

\bibitem[{Tanaka {et~al.}(2002)Tanaka, Takeuchi, \& Ward}]{Tanaka_2002}
Tanaka, H., Takeuchi, T., \& Ward, W.~R. 2002, The Astrophysical Journal, 565, 1257

\bibitem[{Terquem \& Papaloizou(2007)}]{terquem2007}
Terquem, C., \& Papaloizou, J.~C. 2007, The Astrophysical Journal, 654, 1110

\bibitem[{{Van Grootel} {et~al.}(2018){Van Grootel}, {Fernandes}, {Gillon}, {Jehin}, {Manfroid}, {Scuflaire}, {Burgasser}, {Barkaoui}, {Benkhaldoun}, {Burdanov}, {Delrez}, {Demory}, {de Wit}, {Queloz}, \& {Triaud}}]{VanGrootel2018}
{Van Grootel}, V., {Fernandes}, C.~S., {Gillon}, M., {et~al.} 2018, \apj, 853, 30, \dodoi{10.3847/1538-4357/aaa023}

\bibitem[{{Volk} \& {Gladman}(2015)}]{Volk2015ApJ}
{Volk}, K., \& {Gladman}, B. 2015, \apjl, 806, L26, \dodoi{10.1088/2041-8205/806/2/L26}

\bibitem[{{Volk} \& {Malhotra}(2020)}]{Volk2020}
{Volk}, K., \& {Malhotra}, R. 2020, \aj, 160, 98, \dodoi{10.3847/1538-3881/aba0b0}

\bibitem[{Volk \& Malhotra(2024)}]{Volk_2024}
Volk, K., \& Malhotra, R. 2024, The Astronomical Journal, 167, 271

\bibitem[{{Wang} \& {Fischer}(2015)}]{Wang2015}
{Wang}, J., \& {Fischer}, D.~A. 2015, \aj, 149, 14, \dodoi{10.1088/0004-6256/149/1/14}

\bibitem[{{Wang} {et~al.}(2018){Wang}, {Graham}, {Dawson}, {Fabrycky}, {De Rosa}, {Pueyo}, {Konopacky}, {Macintosh}, {Marois}, {Chiang}, {Ammons}, {Arriaga}, {Bailey}, {Barman}, {Bulger}, {Chilcote}, {Cotten}, {Doyon}, {Duch{\^e}ne}, {Esposito}, {Fitzgerald}, {Follette}, {Gerard}, {Goodsell}, {Greenbaum}, {Hibon}, {Hung}, {Ingraham}, {Kalas}, {Larkin}, {Maire}, {Marchis}, {Marley}, {Metchev}, {Millar-Blanchaer}, {Nielsen}, {Oppenheimer}, {Palmer}, {Patience}, {Perrin}, {Poyneer}, {Rajan}, {Rameau}, {Rantakyr{\"o}}, {Ruffio}, {Savransky}, {Schneider}, {Sivaramakrishnan}, {Song}, {Soummer}, {Thomas}, {Wallace}, {Ward-Duong}, {Wiktorowicz}, \& {Wolff}}]{Wang2018}
{Wang}, J.~J., {Graham}, J.~R., {Dawson}, R., {et~al.} 2018, \aj, 156, 192, \dodoi{10.3847/1538-3881/aae150}

\bibitem[{Yuan {et~al.}(2024)Yuan, Gurnis, Asimow, \& Li}]{Yuan2024}
Yuan, Q., Gurnis, M., Asimow, P.~D., \& Li, Y. 2024, Geophysical Research Letters, 51, e2023GL106723

\end{thebibliography}
\bibliographystyle{aasjournal}



\end{document}